\documentclass[12pt]{iopart}

\usepackage{iopams}  
\usepackage{graphicx}
\usepackage{color}

\begin{document}

\title{A positive-definite form of bounce-averaged quasilinear velocity diffusion for the parallel inhomogeneity in a tokamak}

\author{Jungpyo Lee}
\address{MIT Plasma Science and Fusion Center, Cambridge, MA, USA}
\ead{Jungpyo@psfc.mit.edu}

\author{David Smithe}
\address{Tech-X, Boulder, CO, USA}

\author{John Wright}
\address{MIT Plasma Science and Fusion Center, Cambridge, MA, USA}

\author{Paul Bonoli}
\address{MIT Plasma Science and Fusion Center, Cambridge, MA, USA}

\begin{abstract}
In this paper, the analytical form of the quasilinear diffusion coefficients is modified from the Kennel-Engelmann diffusion coefficients to guarantee the positive definiteness of its bounce average in a toroidal geometry. By evaluating the parallel inhomogeneity of plasmas and magnetic fields in the trajectory integral, we can ensure the positive definiteness and help illuminate some non-resonant toroidal effects in the quasilinear diffusion. When the correlation length of the plasma-wave interaction is comparable to the magnetic field variation length, the variation becomes important and the parabolic variation at the outer-midplane, the inner-midplane, and trapping tips can be evaluated by Airy functions. The new form allows the coefficients to include both resonant and non-resonant contributions, and the correlations between the consecutive resonances and in many poloidal periods. The positive-definite form is implemented in a wave code TORIC and we present an example for ITER using this form.
\end{abstract}
\maketitle
\section{Introduction}
In a Fokker-Planck equation, the change of distribution function by RF waves is determined by quasilinear diffusion in velocity space \cite{Stix:AIP1992} (see Appendix A).  The quasilinear theory is sufficiently valid if the perturbation from the background distribution function due to the waves is small enough to be linearized \cite{Cary:PRL1990, laval1999controversies, Lee:POP2011}. The quasilinear diffusion coefficients that were analytically derived by Kennel and Engelmann \cite{Kennel:POF1966} have been used in many numerical codes (e.g. TORIC \cite{Brambilla:PPCF1999} and AORSA \cite{Jaeger:PoP2011}). The Kennel-Engelmann (K-E) coefficients are based on the several assumptions; the particle trajectory is not perturbed by the waves, and on the trajectory the plasma density, temperature and the background magnetic fields are homogenous. Although these assumptions are not satisfied in the toroidal geometry, in which the background magnetic fields vary along the particle trajectory, the K-E coefficients are acceptably applicable because the correlation length of the wave-particle resonance is much shorter than the characteristic length of the variation due to the toroidal geometry \cite{Stix:AIP1992}. Nevertheless, in some conditions as will be shown in this paper, the variation is not negligible compared to the correlation length and the effects of the toroidicity need to be included in the quasilinear formulation. In this paper, we derive an analytic form of the coefficients that can capture the geometry effects.

The toroidal effects on the quasilinear diffusion coefficients have been investigated in many previous studies \cite{Brambilla:PLA1994,smithe:PRL1988,Berry:PoP2016,Bernstein:PF1981,Kerbel:PF1985,Becoulet:PFB1991,Catto:PF1992,Lamalle:PPCF1997,Eester:PPCF2005,Johnson:NF2006,Harvey:RF2001,Kaufman:PF1972,Eriksson:PoP1994,Eriksson:PoP2005}, and they could be summarized by the three types of effects: (1) the change of the resonance correlation length, (2) the additional non-resonant interaction, and (3) the finite particle orbit width. First, the variation of the resonance kernel argument, $\omega-n\Omega-k_\|v_\|$, changes the parallel space dispersion and the correlation between plasmas and waves. Here, $\omega$, $\Omega$, $k_\|$, and $v_\|$ are the wave frequency, the gyrofrequency, the parallel wavenumber, and the parallel velocity, respectively, and $n$ is an integer that determines a type of resonance. In the K-E coefficients, the resonant kernel is represented by a Dirac-delta function with this argument, which can be applicable in the linear variation of the argument. The parabolic variation due to the toroidicity is approximately captured by the modified effective $k_\|$  due to the toroidal broadening \cite{Brambilla:PLA1994} and the modified dispersion function \cite{Brambilla:PLA1994, smithe:PRL1988, Berry:PoP2016}. The more accurate form in the toroidal geometry is obtained by evaluating the trajectory integral \cite{Bernstein:PF1981,Kerbel:PF1985,Becoulet:PFB1991,Catto:PF1992,Lamalle:PPCF1997,Eester:PPCF2005,Johnson:NF2006}. Secondly, the non-resonant interaction when $\omega-n\Omega-k_\|v_\|\neq0$ can contribute to the diffusion significantly if the correlation length is sufficiently long at the outer-midplane or inner-midplane or at the banana tips of the trapped particle \cite{Catto:PF1992,Lamalle:PPCF1997}. The smooth connection between the resonant interactions and the non-resonant interactions is considered in a continuous kernel \cite{Johnson:NF2006}. Finally, the radial departure of the particle trajectory from the magnetic flux surface can change the diffusion coefficients. These finite orbit effects are analytically considered with the modified argument $\omega-n\Omega-k_\|v_\|-\mathbf{k}\cdot \mathbf{v_D}$ \cite{Kaufman:PF1972,Eriksson:PoP1994} where $\mathbf{k}$ is the wavevector and $\mathbf{v_D}$ is the drift velocity, and they are numerically investigated by particle codes \cite{Johnson:NF2006,Harvey:RF2001}. In the new formulation of this paper, the first and second effects are well included, while the finite orbit effects are not considered for simplicity. The finite orbit width effects are important only for the energetic ions.  

A distinctive characteristic of our formulation compared to the K-E formulation is the positive definiteness of its bounce average. The positive definiteness is proven by the symmetric form between the bounce average and the trajectory integral \cite{Bernstein:PF1981,Catto:PF1992,smithe1989local}. The bounce-average of the quasilinear diffusion coefficients is used when taking the bounce-average of the Fokker-Planck equation to eliminate the parallel streaming term \cite{Harvey:IAEA1992}. As will be shown in Section 2, the trajectory integral along the homogenous magnetic field in the K-E coefficients is not consistent with the bounce-average integral in the toroidal geometry, and as a result, the K-E coefficients cannot result in the positive definiteness. The non-positive definite coefficients cause numerical problems as well as unphysical phenomena by violating H-theorem and allowing a growing mode. For the positive-definite form, the trajectory integral needs to be evaluated in the toroidal geometry, as will be done in Section 3. 

For a toroidal axisymmetric geometry, the positive-definiteness of the bounce-averaged quasilinear diffusion coefficients was rigorously proven by Kaufman in \cite{Kaufman:PF1972} using the action-angle variables in terms of the cyclotron motion, the bounce motion, and toroidal motion by the canonical toroidal angular momentum. Because we do not consider radial diffusion in this paper but only the velocity diffusion in the quasilinear diffusion, the diffusion form by Kaufman is not directly applicable. For particle codes, the Monte-Carlo operators were derived for the particle diffusion based on the theory by Kaufman \cite{Eriksson:PoP1994, Eriksson:PoP2005}. In this paper, we only consider the velocity diffusion to be applicable to the coupled codes between Maxwell's equation and the continuum Fokker-Planck equation solvers, while keeping the symmetric form in the bounce-averaged diffusion coefficients.

 We implement the new form of the bounce-averaged diffusion coefficients in a coupled code for ion cyclotron range of frequency (ICRF) waves, TORIC-CQL3D \cite{Lee:PoP2017}. Although the new form can be applicable to all types of resonances, ICRF waves can have more significant toroidal effects in the form due to the varying gyrofrequency along the trajectory. As an example, the minority fundamental ion cyclotron damping in ITER is examined to evaluate the diffusion coefficients. The results are compared with the original K-E coefficients, as will be shown in Section 4.
 
Several important features of the diffusion coefficients are found in our formulation. In the trajectory integral, the correlation between consecutive resonances can be included and it is important if two consecutive resonances are located around the specific poloidal locations, where the correlation length is comparable to the variation length. In this location, the non-resonant contribution to the velocity diffusion may also not be negligible. Because the trajectory integral does not guarantee the periodicity of the phase in each poloidal period, the correlations between resonances in a couple of periods remain and make the average diffusion coefficient different in terms of the number of evaluation periods  \cite{Harvey:RF2001}. These features will be explained in Section 4.

The rest of the paper is organized as follows. In Section \ref{sec:2}, we summarize the bounce-average of the K-E diffusion coefficients and examine the problems with its use in the toroidal geometry. In Section \ref{sec:3}, we suggest a modified form of the bounce-average coefficients that yield symmetric and positive-definite properties in the toroidal geometry. The implementation of the modified coefficients in TORIC  \cite{Brambilla:PPCF1999} is also explained. In Section \ref{sec:4}, three features of the coefficients are discussed by investigating an example. Finally, a discussion is given in Section \ref{sec:5}. 


\section{Bounce-average of K-E coefficients}\label{sec:2}
 The gyro-averaged quasilinear diffusion coefficients $D_{ql}$ in the homogeneous plasma and magnetic fields can be derived by linearizing the Vlasov equation with electromagnetic waves \cite{Stix:AIP1992,Kennel:POF1966,smithe:PRL1988}, as summarized in Appendix B. Using the Fourier analysis on the electric fields for Eqs. (\ref{Dql1app}) and (\ref{delta2}), the Kennel-Engelmann quasilinear diffusion coefficient tensor is
\begin{eqnarray}
 \fl D_{ql}     
&\simeq&    \frac{ q^2}{m^2} Re\bigg\{\sum_n\int d\mathbf{k}_1\int d\mathbf{k}_2   \left( (\mathbf{P_n}(\mathbf{k}_2)\cdot\mathbf{E(\mathbf{k}_2)})\mathbf{G}(\mathbf{k}_2))^* e^{i(\mathbf{k}_{1}-\mathbf{k}_{2}) \cdot \mathbf{r}(t)}\right) \nonumber \\
\fl && \times \int_{-\infty}^t dt^{\prime} \left( (\mathbf{P_n}(\mathbf{k}_1) \cdot\mathbf{E(\mathbf{k}_1)})\mathbf{G}(\mathbf{k}_1)\right)  e^{i\int_{t\prime}^t dt^{\prime\prime}(\omega-n\Omega-k_{\|1}v_\|)}\bigg\} ,  \label{Dql1}
 \end{eqnarray}
where $\mathbf{E}$ is the electric field, $\mathbf{r}$ is the space vector, and the superscript $*$ denotes the conjugate transpose. As shown in Appendix B,  the gyro-average results in the polarization vector $\mathbf{P_n}$, and the diffusion vector $\mathbf{G}$  \cite{Stix:AIP1992,Kennel:POF1966}, satisfying  
  \begin{eqnarray}
  \mathbf{P_n} \cdot \mathbf{E}&=&E_{\mathbf{k} ,+} \frac{J_{n-1}}{\sqrt{2}} +E_{\mathbf{k} ,-} \frac{J_{n+1}}{\sqrt{2}}+\frac{v_{\|}}{v_{\perp}}E_{\mathbf{k} ,\|} J_{n} ,\label{Pn_text}\\
  \mathbf{G}&=&\left(1-\frac{k_{\|}v_{\|}}{\omega}\right)\hat{\mathbf{v}}_\perp+\frac{k_{\|}v_{\perp}}{\omega}\hat{\mathbf{v}}_\|, \label{G_text}
   \end{eqnarray}
   where $J_{n}=J_n(k_\perp \rho_i)$ is the Bessel function of the first kind for the order n, $v_\perp$ is the perpendicular velocity, $\hat{\mathbf{v}}_\perp$, $\hat{\mathbf{v}}_\|$ are the unit vectors in the velocity space along the perpendicular and the parallel direction, respectively, $\rho_i=v_\perp/\Omega$ is the ion Larmor radius, $k_\perp$ is the perpendicular wavevector, $E_{\mathbf{k} ,+}$, and $E_{\mathbf{k} ,-}$ are the left-hand and the right hand polarized electric fields, respectively, and $E_{\mathbf{k} ,\|}$ is the parallel electric field  at the local coordinate of the current position with $t$. Here, Eq. (\ref{Dql1}) includes the interference between the different spectral modes $\mathbf{k}_1$ and $\mathbf{k}_2$ unlike the original K-E coefficient in Eq. (\ref{Dql1app}) because in the toroidal geometry the inhomogeneous plasmas and magnetic fields results in the interference \cite{Jaeger:NF2006}.
   
Taking a bounce-average on the quasilinear diffusion term $Q$ in Eq. (\ref{Q1}) induces the bounce-averaged diffusion coefficients $\langle D_{ql} \rangle_b$,
\begin{eqnarray}
 \fl &&\langle Q \rangle_b=\langle \nabla_\mathbf{v} \cdot  D_{ql}   \cdot \nabla_\mathbf{v} f(\mathbf{v})\rangle_b =\frac{1}{\lambda_p}\nabla_\mathbf{v_c} \cdot \lambda_p \left \langle \frac{\partial \mathbf{v_c}}{\partial \mathbf{v}} \cdot D_{ql} \cdot \frac{\partial \mathbf{v_c}}{\partial \mathbf{v}}^T \right \rangle_b  \cdot \nabla_\mathbf{v_c} f (\mathbf{v_c}),  \label{Dqlb1}
  \end{eqnarray}
  where  $\langle x \rangle_b=(1/T_p)\int_0^{l_p} dl(x/v_\|) =(1/T_p) \int_0^{T_p} dt  \,x$ is the bounce average, $l$ is the distance and $t$ is the time along the gyro-averaged orbit trajectory, respectively, $l_p$ is the distance of one bounce orbit and $T_p=\int_0^{l_p} (dl/v_\|)$ is the bounce time. Here, $\mathbf{v_c}$ is the invariant velocity space coordinate (e.g. $v$ and $\mu$ in Appendix C or the velocity at the outer-midplane for CQL3D \cite{Harvey:IAEA1992}), while $\mathbf{v}$ is the velocity space coordinate with the varying components in the trajectory time (e.g. $v_\|$ and $v_\perp$ in the toroidal geometry) that are used to define the diffusion tensor $D_{ql}$ as in Eqs. (\ref{Pn_text}) and (\ref{G_text}). To transform the coordinate, the Jacobian tensor between two coordinates ${\partial \mathbf{v_c}}/{\partial \mathbf{v}}$ is used and  $\lambda_p (\mathbf{v_c})=v_{\|0}T_p$ is used to conserve the total number of particles in a flux tube \cite{Harvey:IAEA1992}.
  
 The bounce-average of the quasilinear diffusion coefficient in Eq. (\ref{Dqlb1}) is
\begin{eqnarray}
 \fl&& \left \langle \frac{\partial \mathbf{v_c}}{\partial \mathbf{v}} \cdot D_{ql} \cdot \frac{\partial \mathbf{v_c}}{\partial \mathbf{v}}^T \right \rangle_b    \nonumber \\ 
 \fl&\simeq&     \frac{q^2}{T_pm^2}  Re\bigg\{\sum_n\int d\mathbf{k}_1\int d\mathbf{k}_2 \int_0^{T_p} dt  \frac{\partial \mathbf{v_c}}{\partial \mathbf{v}} \cdot ((\mathbf{P_n}(\mathbf{k}_2)\cdot\mathbf{E(\mathbf{k}_2)})\mathbf{G}(\mathbf{k}_2))^* e^{i(\mathbf{k}_{1}-\mathbf{k}_{2}) \cdot \mathbf{r}(t)} \nonumber \\
\fl && \times \int_{-\infty}^t dt^{\prime} \left( (\mathbf{P_n}(\mathbf{k}_1) \cdot\mathbf{E(\mathbf{k}_1)})\mathbf{G}(\mathbf{k}_1)\right)\cdot \frac{\partial \mathbf{v_c}}{\partial \mathbf{v}}^T e^{i\int_{t\prime}^t dt^{\prime\prime}(\omega-n\Omega-k_{\|1}v_\|)}\bigg\}.  \label{Dqlb2}
 \end{eqnarray}
  For the K-E coefficients, the assumption of the homogenous plasmas and magnetic fields results in the constant phase $\omega-n\Omega-k_\|v_\|$ along the trajectory, and the phase integral $\int_{t^\prime}^t dt^{\prime\prime}(\omega-n\Omega-k_{\|}v_\|)=(\omega-n\Omega-k_{\|}v_\|)(t-{t^\prime})$ results in the Dirac-delta function in $t^\prime$. Thus, the bounce-average of the K-E coefficient is
 \begin{eqnarray}
\fl && \left \langle \frac{\partial \mathbf{v_c}}{\partial \mathbf{v}} \cdot D_{ql} \cdot \frac{\partial \mathbf{v_c}}{\partial \mathbf{v}}^T \right \rangle_b \nonumber \\
\fl&=&      \frac{\pi q^2}{T_p m^2}  Re\bigg\{\sum_n\int d\mathbf{k}_1\int d\mathbf{k}_2 \int_0^{T_p} dt  \frac{\partial \mathbf{v_c}}{\partial \mathbf{v}} \cdot (\mathbf{P_n(\mathbf{k}_2)}\cdot\mathbf{E(\mathbf{k}_2)}) ^*(\mathbf{G}(\mathbf{k}_2)^* \mathbf{G}(\mathbf{k}_1)) \nonumber \\
\fl &&\times  (\mathbf{P_n}(\mathbf{k}_1)\cdot\mathbf{E(\mathbf{k}_1)}) \cdot \frac{\partial \mathbf{v_c}}{\partial \mathbf{v}}^T 
e^{i(\mathbf{k}_{1}-\mathbf{k}_{2}) \cdot \mathbf{r}(t)}\delta(\omega-n\Omega-k_{\|1}v_\|)   \bigg\} \label{Dql2},
 \end{eqnarray}
 where $\delta$ represents the Dirac-delta function. The bounce-averaged coefficients in Eq. (\ref{Dql2}) is widely used in many codes that couples the Maxwell equation solver with the continuum Fokker-Planck codes (e.g. AORSA-CQL3D \cite{Jaeger:PoP2011, Jaeger:NF2006} and TORIC-CQL3D \cite{Lee:PoP2017}) 

For a toroidal geometry, Eq. (\ref{Dql2}) is used by changing the argument of the delta function in terms the integral variable $t$ (i.e. $\omega-n\Omega(t)-k_{\|1}(t)v_\|(t)$). However, the symmetry in $\mathbf{k}_1$ and $\mathbf{k}_2$ is broken in Eq. (\ref{Dql2}), because the Dirac-delta function depends on only one of wavevectors such as $\mathbf{k}_1$ but the poloidal spectral modes are coupled because of the variation of $\mathbf{P_n}$ and $\mathbf{G}$ in terms of $t$. As a result, the positive definiteness cannot be guaranteed. Instead, it is likely to be partially non-positive due to the interference between the different modes given by the term, $\exp({i(\mathbf{k}_{2}-\mathbf{k}_{1}) \cdot \mathbf{r}(t)})$. The negative values of bounce-averaged quasilinear diffusion generate a growing mode that is numerically unstable and non-physical because it violates the H-theorem \cite{Kennel:POF1966}. Eliminating these negative values from the coefficients therefore leads to an error in their evaluation. 
 
 Figure 1 shows the error of the radial power absorption profiles due to the negative values of bounce-averaged diffusion. As shown in the gap between the graphs, the error due to the negative $\langle D_{ql} \rangle_b$ becomes significant, as its velocity grid resolution increases. This may be the case because a small number of grid points is likely better to allow the average of the negative values to be positive in a large grid spacing. As the resolution is made finer, more power resides in regions of negative quasilinear diffusion. This can be problematic for the numerical convergence of a code. In Figure 1, The error vanishes for a small radius (as $r/a\rightarrow 0$), implying that the negative values are due to the effect of the toroidal geometry with the finite aspect ratio. Although this problem may be reduced somewhat by using a coarse velocity grid, the positive-definite form of the bounce-averaged quasilinear diffusion is required to ensure the accuracy of the evaluation. The error due to the negative diffusion in Figure 1 will be eliminated by the positive-definite form in this paper, as will be shown in Figure 5. 
 
 
     \begin{figure} 
\includegraphics[scale=0.5]{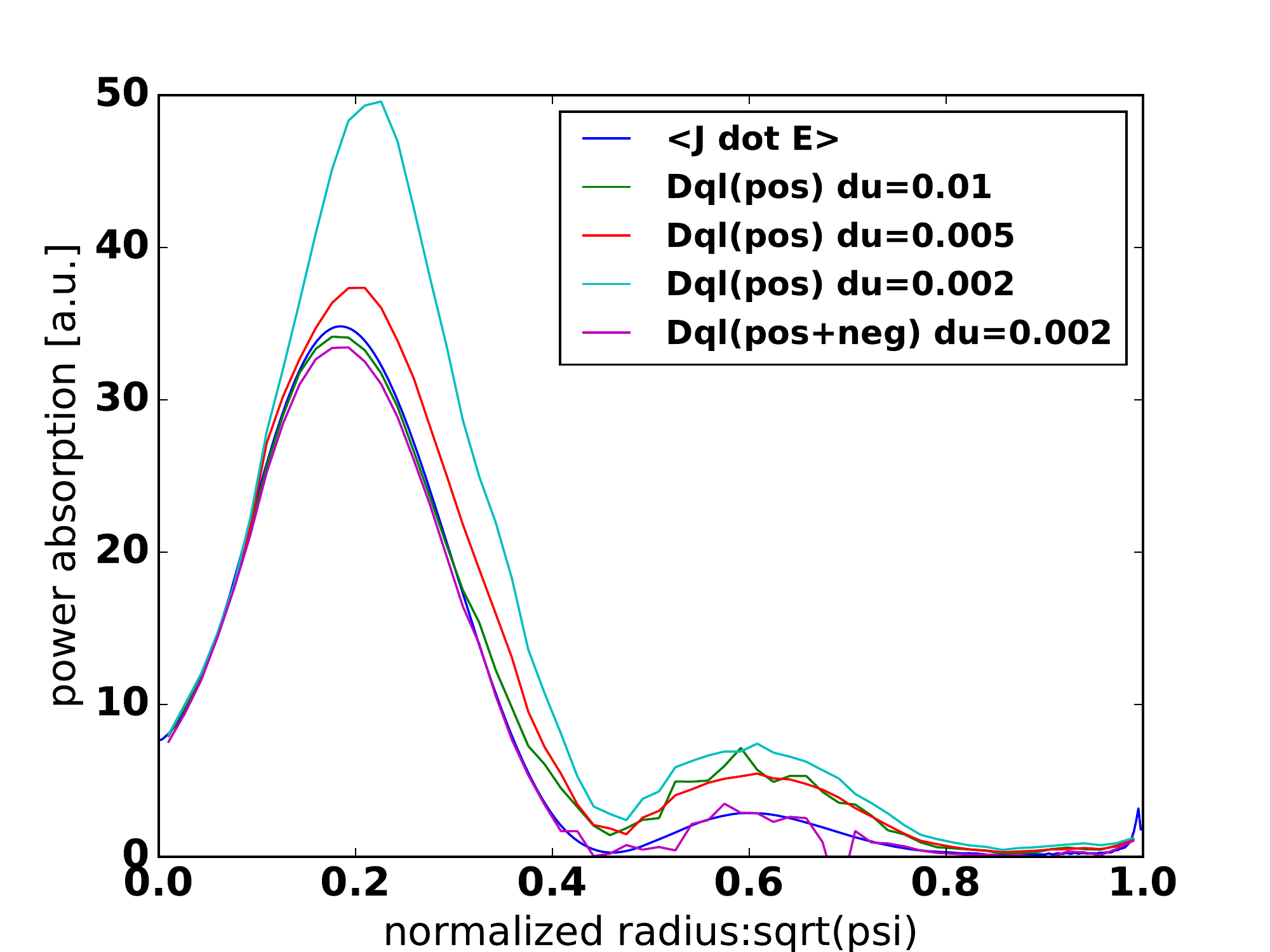}
\caption{Radial profiles of the ICRF power absorption for Maxwellian plasmas using the flux surface average of $\mathbf{J} \cdot \mathbf{E}$ (blue) and $\dot{W}=\int d\mathbf{v}  (mv^2/2)(d/dv_c)(\langle D_{ql} \rangle_b df_M/dv)$ with the Kennel-Engelmann coefficients in different velocity space grid resolutions: the normalized grid spacing $\Delta v= 0.01$ (green), $\Delta v= 0.005$ (red), and $\Delta v= 0.002$ (cyan) only when the positive values of $\langle D_{ql} \rangle_b$ are included in the evaluation. The violet graph is obtained with $\Delta v= 0.002$ when both positive and negative values $\langle D_{ql} \rangle_b$ are included.}
\label{fig:d_opt}
\end{figure}

\section{Symmetric form for a toroidal axisymmetric geometry}\label{sec:3}
\subsection{Modification to the symmetric form}\label{sec:3-1}
 In this section, we derive a symmetric form of the bounce-averaged quasilinear diffusion coefficients based on the proof in Appendix B of \cite{smithe1989local} for the inhomogeneous plasmas and magnetic fields in a toroidal axisymmetric geometry.

We recall that the broken symmetry in Eq. (\ref{Dql2}) is because the variation of the resonance argument $``\omega-n\Omega-k_\|v_\|"$ is not applied equivalently to the integrals of $t$ and $t^\prime$, and as a result the Dirac-delta function does not depend on $k_{\|2}$ but only on $k_{\|1}$. By keeping the toroidal variation in both integrals in $t$ and $t^\prime$ equivalently, we will show the symmetric form in Eq. (\ref{Dql3}).

\subsubsection{Bounce integral}

The bounce-averaged diffusion can be derived by integrating the quasilinear equation along an unperturbed orbit from $t_\infty$ to $t$, and dividing that result by ($t-t_\infty$) to get the average, 
\begin{eqnarray}
\fl \lim_{t_\infty\rightarrow -\infty} \frac{\int_{t_\infty}^t dt^{\prime\prime} \frac{\partial f_0}{\partial t}}{t-t_\infty}= \lim_{t_\infty\rightarrow -\infty} \frac{ f_0(\mathbf{r},\mathbf{v},t)- f_0(\mathbf{r_\infty},\mathbf{v_\infty},t_\infty)}{t-t_\infty}\nonumber \\
\fl=-\lim_{t_\infty\rightarrow -\infty} \frac{1}{t-t_\infty}\int_{t_\infty}^t dt^{\prime\prime} \bigg[\frac{q}{m}\bigg(\nabla_\mathbf{v_c} \cdot \frac{\partial \mathbf{v_c}}{\partial \mathbf{v}}(\mathbf{r_0}(t^{\prime\prime}),\mathbf{v_0}(t^{\prime\prime})) 
\cdot \left(\mathbf{E}(\mathbf{r_0}(t^{\prime\prime}))+ \mathbf{v_0}(t^{\prime\prime})\times\mathbf{B}(\mathbf{r_0}(t^{\prime\prime})\right)^*\nonumber \\
\fl +\nabla_\mathbf{x_c} \cdot \frac{\partial \mathbf{x_c}}{\partial \mathbf{v}}(\mathbf{r_0}(t^{\prime\prime}),\mathbf{v_0}(t^{\prime\prime}))\cdot \left(\mathbf{E}(\mathbf{r_0}(t^{\prime\prime}))+ \mathbf{v_0}(t^{\prime\prime})\times\mathbf{B}(\mathbf{r_0}(t^{\prime\prime})\right)^*\bigg)f_1(\mathbf{r_0}(t^{\prime\prime}),\mathbf{v_0}(t^{\prime\prime}))\bigg] \label{proof1},
\end{eqnarray}
where Eq. (\ref{AppGam}) is used with the constants-of-motions $\mathbf{v_c}$ and its conjugate coordinate $\mathbf{x_c}$. Here, $f_0$ is the background distribution function varying in a slow time scale, $f_1$ is the perturbed distribution function due to the RF waves, and $\mathbf{r_0}(t^{\prime\prime})$ and $\mathbf{v_0}(t^{\prime\prime})$ are the vectors of $\mathbf{r}$ and $\mathbf{v}$, respectively, along the unperturbed orbits at the time $t^{\prime\prime}$.

The outer derivative term in $\nabla_\mathbf{x_c}$ of Eq. (\ref{proof1}) can be eliminated by the integrate over the three conjugate coordinate $\mathbf{x_c}$. Here it is easiest to assume the favorite constants-of-motion for tokamaks, $\mathbf{v_c}=\{E,\mu,P_\varphi\}$, e.g., energy, magnetic moment, and canonical angular momentum, with $\mathbf{x_c}=\{t,\phi,\varphi_c\}$ as the conjugate coordinates. 
The time integral we have already done is the first of the three integrals, since time is the conjugate coordinate to the energy constant-of-motion.  For this, we make the slight change that instead of taking the limit of $t_\infty \rightarrow -\infty$, we take the limit of
\begin{eqnarray}
t_\infty=-N_b T_p(\mathbf{v_c}),
\end{eqnarray}
where $N_b$ is the number of the bounce period. The other two integrals over the conjugate coordinates are just two angle coordinate averages over $2\pi$ radians, e.g., $(2\pi)^{-2} \int d\varphi_c d\phi $. With $t_\infty$ an integer number of bounce periods, the integral of the coordinate derivatives,
\begin{eqnarray}
 \int d\varphi_c\int d\phi \int^t_{t_\infty}  dt^{\prime\prime} \nabla_\mathbf{x_c} \cdot A=0
\end{eqnarray}
goes to zero for an arbitrary $A$, since each is an integral of the derivative of a periodic quantity, integrated over an integer number of periods.  

Then, taking the average of Eq. (\ref{proof1}) results in the symmetric form,

\begin{eqnarray}
\fl \lim_{N_b\rightarrow \infty} \frac{1}{(2\pi)^2} \int d\varphi_c\int d\phi \frac{ f_0(\mathbf{r},\mathbf{v},t)- f_0(\mathbf{r_\infty},\mathbf{v_\infty},t_\infty)}{t-t_\infty}\nonumber \\
\fl=\nabla_\mathbf{v_c} \cdot \lim_{N_b\rightarrow \infty}\frac{1}{t-t_\infty}  \frac{1}{(2\pi)^2}\int d\varphi_c\int d\phi \int_{t_\infty}^t dt^{\prime\prime} \bigg[\frac{q^2}{m^2}\frac{\partial \mathbf{v_c}}{\partial \mathbf{v}}(\mathbf{r_0}(t^{\prime\prime}),\mathbf{v_0}(t^{\prime\prime})) \cdot \left(\mathbf{E}(\mathbf{r_0}(t^{\prime\prime}))+ \mathbf{v_0}(t^{\prime\prime})\times\mathbf{B}(\mathbf{r_0}(t^{\prime\prime})\right)^* \nonumber \\
\fl  \int_{t_\infty}^{t^{\prime\prime}} dt^{\prime} \left(\mathbf{E}(\mathbf{r_0}(t^{\prime}))+ \mathbf{v_0}(t^{\prime})\times\mathbf{B}(\mathbf{r_0}(t^{\prime})\right) \cdot \frac{\partial \mathbf{v_c}}{\partial \mathbf{v}}^T(\mathbf{r_0}(t^{\prime}),\mathbf{v_0}(t^{\prime})) \bigg] \cdot \nabla_\mathbf{v_c} f_0(\mathbf{v_c}), \label{proof2}
\end{eqnarray} 
where $f_1$ in Eq. (\ref{Appf1}) is replaced and the outer $\nabla_\mathbf{v_c}$ derivative is then let to slide to the outside of the integrals, since the integrals do not affect the constants of motion. In computational practice, we do not take the limit $N_b\rightarrow \infty$, but instead take $N_b \rightarrow N_{decor}$, where $N_{decor}$ is a number of bounces such that decorrelation effects make $f_1$ independent of $t_\infty$ when it exceeds $t_d=-N_{decor} T_p$. For most particles in a tokamak, a single bounce period suffices, but it may depends on the decorrelation mechanism as will be mentioned in Section 4.3. 

We also make the quasilinear assumption, e.g., that the changes to the unperturbed distribution function are small over the decorrelation time, $t_d$.  That is
\begin{eqnarray}
\fl \frac{\partial f_0}{\partial t} \simeq \frac{1}{(2\pi)^2} \int d\varphi_c\int d\phi \frac{ f_0(\mathbf{r},\mathbf{v},t)- f_0(\mathbf{r_d},\mathbf{v_d},t_d)}{t-t_d} \equiv \nabla_\mathbf{v_c} \cdot \mathcal{D} (\mathbf{v_c}) \cdot \mathbf{v_c} f_0(\mathbf{v_c}),
\end{eqnarray} 
where the diffusion coefficient is
\begin{eqnarray}
\fl \mathcal{D} (\mathbf{v_c}) \equiv  \frac{q^2}{m^2}\frac{1}{(2\pi)^2} \frac{1}{t-t_d} \int d\varphi_c\int d\phi \int_{t_\infty}^t dt^{\prime\prime}\int_{t_\infty}^{t^{\prime\prime}} dt^{\prime}  \mathcal{I} (\mathbf{r} ,\mathbf{v}, t^\prime, t^{\prime\prime}  ).
\end{eqnarray} 
Here, the integrand of the $t^\prime$ and $t^{\prime\prime}$ integrals in $\mathcal{I}$ is symmetric by Eq. (\ref{proof2}), giving
\begin{eqnarray}
\mathcal{I} (\mathbf{r} ,\mathbf{v}, t^\prime, t^{\prime\prime}) =\mathcal{I} (\mathbf{r} ,\mathbf{v},  t^{\prime\prime},t^\prime).  \label{tex1}
\end{eqnarray} 
 However, the integration limits of the $t^\prime$ and $t^{\prime\prime}$ integrals are not the same, and so we do not yet have obvious symmetry. 

\subsubsection{Hidden symmetry of the time integral limits}
The integrand limits of the double time integral are for the upper triangle of a symmetric rectangular domain in $t^\prime-t^{\prime\prime}$  Thus, the integrals are half the integral over the rectangular domain, giving

\begin{eqnarray}
\int_{t_d}^t dt^{\prime\prime}\int_{t_d}^{t^{\prime\prime}} dt^{\prime}  \mathcal{I} (\mathbf{r} ,\mathbf{v}, t^\prime, t^{\prime\prime}) =\frac{1}{2}\int_{t_d}^t dt^{\prime\prime}\int_{t_d}^{t} dt^{\prime}  \mathcal{I} (\mathbf{r} ,\mathbf{v},  t^{\prime\prime},t^\prime)  \label{tex2},
\end{eqnarray} 
where the symmetry of the integrand in Eq. (\ref{tex1}) is used. With the rectangular domain integral limits, we finally achieve perfect Hermitian symmetry, and have
\begin{eqnarray}
\mathcal{D} (\mathbf{v_c}) \equiv  \frac{q^2}{2m^2}\frac{1}{(2\pi)^2} \frac{1}{t-t_d} \int d\varphi_c\int d\phi \mathcal{A}^{\forall} (\mathbf{r},\mathbf{v}) \mathcal{A}(\mathbf{r},\mathbf{v})\label{proof3}
\end{eqnarray} 
where the vector is
\begin{eqnarray}
 \mathcal{A} (\mathbf{r},\mathbf{v}) = \int_{t_d}^{t} dt^{\prime} \left(\mathbf{E}(\mathbf{r_0}(t^{\prime}))+ \mathbf{v_0}(t^{\prime})\times\mathbf{B}(\mathbf{r_0}(t^{\prime})\right) \cdot \frac{\partial \mathbf{v_c}}{\partial \mathbf{v}}^T(\mathbf{r_0}(t^{\prime}),\mathbf{v_0}(t^{\prime})) . 
\end{eqnarray} 

\subsubsection{Approximation of the symmetric form}
 Using the Fourier analyzed fluctuating electric field, $\mathbf{E}=\sum_\mathbf{k}  \mathbf{E}_\mathbf{k}  \exp(i\mathbf{k} \cdot \mathbf{r}-i \omega_\mathbf{k} t)$ and Faraday's law, we can simplify the symmetric form in Eq. (\ref{proof3}). As done in the derivation for K-E coefficients in Eq. (\ref {f_k1}), 
\begin{eqnarray}
\fl && \mathcal{A}(\mathbf{r},\mathbf{v}) = \int_0^{t-t_d} d\tau e^{i\int_0^\tau d \tau^{\prime} (\omega-k_\|v_\|)-i\lambda \sin(\eta+\int_0^\tau d  \tau^{\prime}  \Omega )}  \bigg\{ \cos{(\eta+\int_0^\tau d  \tau^{\prime} \Omega )}(( E_{\mathbf{k} ,+}+E_{\mathbf{k} ,-}) \mathbf{G}-E_{\mathbf{k},\|}  \mathbf{H})\nonumber\\
\fl &&-i\sin{(\eta+\Omega \int_0^\tau d  \tau^{\prime})}( E_{\mathbf{k} ,+}-E_{\mathbf{k} ,-}) \mathbf{G} +E_{\mathbf{k} ,\|} \hat{\mathbf{v}}_\|\bigg \} \cdot \frac{\partial \mathbf{v_c}}{\partial \mathbf{v}}^T, \label{sym_A2}
\end{eqnarray} 
where $\eta=\phi-\beta$, and the diffusion vector $ \mathbf{H}$ is defined from $V$ in Eq. (\ref {f_k1}), as defined for  $ \mathbf{G}$ from $U$, giving
 \begin{eqnarray}
   \mathbf{H}&=&\frac{k_{\perp} }{\omega}\left(v_{\perp}\hat{\mathbf{v}}_\|-v_{\|}\hat{\mathbf{v}}_\perp\right).\label{KE_H1}
   \end{eqnarray}
  Using the Bessel function expansion, Eq. (\ref{sym_A2}) is
  \begin{eqnarray}
\fl && \mathcal{A} (\mathbf{r},\mathbf{v}) = \sum_n \int_0^{t-t_d} d\tau e^{i\int_0^\tau d  \tau^{\prime}(\omega-n\Omega -k_\|v_\|)-in\eta}  \bigg\{ \frac{n}{\lambda}J_n(( E_{\mathbf{k} ,+}+E_{\mathbf{k} ,-}) \mathbf{G}-E_{\mathbf{k},\|}  \mathbf{H})\nonumber\\
\fl &&-\frac{dJ_n}{d\lambda}( E_{\mathbf{k} ,+}-E_{\mathbf{k} ,-}) \mathbf{G} +J_n E_{\mathbf{k} ,\|} \hat{\mathbf{v}}_\|\bigg \} \cdot \frac{\partial \mathbf{v_c}}{\partial \mathbf{v}}^T \label{sym_A30}\\
\fl && \simeq \sum_n \int_0^{\infty} d\tau e^{i\int_0^\tau d  \tau^{\prime}(\omega-n\Omega -k_\|v_\|)-in\eta}\left(\mathbf{G}(\mathbf{k}) \cdot \frac{\partial \mathbf{v_c}}{\partial \mathbf{v}}^T \right)(\mathbf{P_n}\cdot\mathbf{E(\mathbf{k})}),\label{sym_A3}
\end{eqnarray} 
where Bessel functions identifies and Eq. (\ref{appG2}) are used from Eq. (\ref{sym_A30}) to Eq. (\ref{sym_A3}), 
  \begin{eqnarray}
   -\frac{n}{\lambda}\mathbf{H}+ \hat{\mathbf{v}}_\|&=&  \frac{v_\|}{v_\perp} \mathbf{G} +\frac{\omega-n\Omega-k_\|v_\|}{\omega} \left(\hat{\mathbf{v}}_\|-\frac{v_{\|}}{v_{\perp}}\hat{\mathbf{v}}_\perp\right)\simeq  \frac{v_\|}{v_\perp} \mathbf{G}.\label{appG2}
   \end{eqnarray}
   The last approximation in Eq. (\ref{appG2}) is acceptable because of $\omega-n\Omega-k_\|v_\|\simeq 0$ for the region of the interest (even in the non-resonant contribution). Each vector of $\mathcal{A}$ and $\mathcal{A}^{\forall}$ has the summation over $n$ as a result of Bessel function expansion, but it can be reduced to one summation by the average over gyroangle $\phi$,
        \begin{eqnarray}
 \sum_n\sum_{n^\prime} \int d\phi e^{in(\phi-\beta)}e^{-in^\prime(\phi-\beta)} \rightarrow  \sum_n, 
   \end{eqnarray}
 because there is a contribution only when $n^\prime=n$. 
 
 Thus, the final symmetric form for this paper is
\begin{eqnarray}
\mathcal{D} (\mathbf{v_c})
&\simeq & \left \langle \frac{\partial \mathbf{v_c}}{\partial \mathbf{v}} \cdot D_{ql} \cdot \frac{\partial \mathbf{v_c}}{\partial \mathbf{v}}^T \right \rangle_b = \frac{q^2}{2 m^2 t_d}\sum_n \mathbf{T_n}^*\mathbf{T_n},  \label{Dql3} \\
&& \mathbf{T_n} \equiv \int d\mathbf{k} \int^{t_d}_{0}dt \left(\mathbf{G}(\mathbf{k}) \cdot \frac{\partial \mathbf{v_c}}{\partial \mathbf{v}}^T \right)(\mathbf{P_n}\cdot\mathbf{E(\mathbf{k})}) e^{-i\Phi_n(t,k_{\|})} \label{Dql4},
\end{eqnarray}
where the phase integral is 
\begin{eqnarray}
\Phi_n(t,k_{\|})=\int_{0}^t dt^{\prime\prime}(\omega-n\Omega-k_{\|}v_\|). 
\end{eqnarray}
 Using $\mathbf{r}(t=0)=0$ as the reference position of the Fourier analysis for $\mathbf{E}(\mathbf{k})$, it results in $e^{i\mathbf{k}\cdot \mathbf{r}(t=0)}=1$ and there is no explicit interference phase by $e^{i((\mathbf{k}_1-\mathbf{k}_2)\cdot \mathbf{r})}$ of Eq. (\ref{Dql1}) in this form. Instead, the phase change by $\mathbf{k}\cdot \mathbf{r}(t)=\int_{0}^{t} k_\|v_\| dt$ in the parallel motion is included in $\Phi(t,k_\|)$. The interference between $\Phi(t,k_{\|2})$ and $\Phi(t^\prime,k_{\|1})$ of the trajectory integral and bounce integral is implicitly determined by the difference between $\Phi(t,k_\|)$ of each $\mathbf{T_n}$ in Eq. (\ref{Dql3}).   
 
Because we ignore the radial departure from the flux surface due to the finite orbit width for the convenience, the average over the cannonical toroidal angle $\int d\varphi_c$ in Eq. (\ref{proof3}) is replaced by the average over the real toroidal angle $\int_0^{2\pi} d\varphi$. Ignoring the finite orbit width effect may result in the error of the symmetric form, but it may be acceptable because the radial departure from the flux surface is much smaller than the major radius to define the toroidal angle.


  \subsection{Trajectory integral}
  The trajectory integral in terms of $t$ in Eq. (\ref{Dql4}) can be analytically evaluated by the stationary approximation using the expansion of $t$ around the resonance points \cite{Bernstein:PF1981,Kerbel:PF1985,Becoulet:PFB1991,Catto:PF1992,Lamalle:PPCF1997,Eester:PPCF2005, Johnson:NF2006}. The contribution of the phase integral to the $\langle D_{ql}\rangle_b$ is significant around the resonance when $d\Phi_n/dt=\omega-n\Omega-k_{\|}v_\|=0$. In addition to the resonance contribution, we also include the contribution of the phase integral by the non-resonant interaction if the correlation length of the interaction is comparable to the length scale of the magnetic field variation, as also suggested in \cite{Catto:PF1992,Lamalle:PPCF1997}.
  
 For the resonant contribution, the phase around the resonance points can be approximated using the Taylor series up to third order, as done in \cite{Bernstein:PF1981,Kerbel:PF1985,Becoulet:PFB1991,Catto:PF1992,Lamalle:PPCF1997,Eester:PPCF2005,Johnson:NF2006}, giving 
     \begin{eqnarray}
  \Phi_n(t,k_{\|})\simeq  \Phi_n(t_{r},k_{\|})+\frac{1}{2}\frac{d^2 \Phi_n}{dt^2}\bigg|_{t=t_{r}}(t-t_{r})^2+\frac{1}{6}\frac{d^3 \Phi_n}{dt^3}\bigg|_{t=t_{r}}(t-t_{r})^3, \label{phasesmallt}
  \end{eqnarray}
  where the $t_r$ indicates $t$ at the resonance location of $d\Phi_n/dt(t=t_r)=0$. The range of integral in terms of $t$ in Eq. (\ref{Dql4}) can be reduced to the bounce time period $T_p$, assuming the correlation between each period is negligible. The violation of this assumption that results in the non-periodicity of the coefficients will be discussed in Sec. 4.3. For convenience of the description, we divide the full range into two separate ranges, the trajectory in the upper-midplane and that in the lower-midplane, as shown in Figure 2. For up-down symmetric tokamak, the phase integrals in the two ranges are symmetric. This separation of the range results in the continuous connection between the resonance contribution and the non-resonant contribution at the outer-midplane or at the inner-midplane, as will be shown in Figure 3.    
  
      \begin{figure} 
    \includegraphics[scale=0.47]{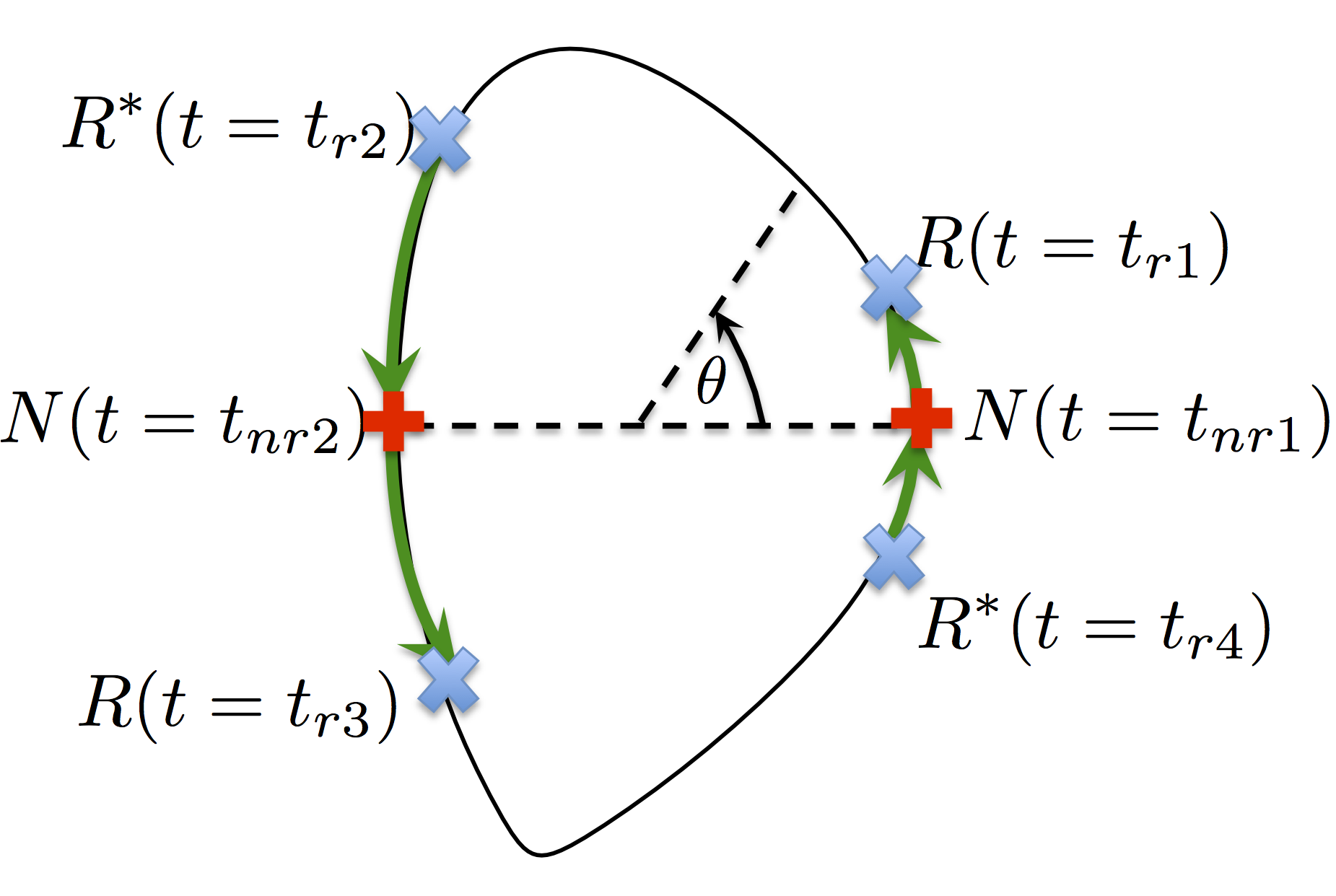}
\caption{ Schematic of the effective trajectory integral evaluation in a poloidal cross section for a flux surface. The horizontal dashed line represents the midplane. The blue cross markers ``x" represent the resonance location $t=t_r$ satisfying $d\Phi_n/dt=0$ and evaluating the integral by $R(t_r)$. Each path of the right arrows corresponds to the trajectory integral range that is taken into account in $R(t_r)$ for each resonance pint. The red plus markers ``+" represent the non-resonance location $t=t_{nr}$ satisfying $d^2\Phi_n/dt^2=0$ for passing particles and evaluating the integral by $N(t_r)$. 
}
\label{fig:d_opt}
\end{figure}
The integral of the phase in Eq. (\ref{phasesmallt}) can be evaluated by the incomplete Airy function. For convenience, we derive the integral for $\mathbf{T_n}^*$ in this section. For the positive parallel velocity particles in the upper-midplane, it results in  
          \begin{eqnarray}
\fl  && \int^{-T_{p/2}}_{-T_{p}} dt g(t) e^{i\Phi_n(t,k_{\|})}\simeq   \int^{T_{p/2}}_{0} dt g(t) e^{i\Phi_n(t,k_{\|})} \nonumber\\
\fl &\simeq&g(t=t_r) e^{ i\Phi_n(t_{r},k_{\|})} \int^{T_{p/2}}_0 dt  e^{i (1/2)({d^2\Phi_n}/{dt^2})|_{t=t_{r}}(t-t_{r})^2+i({1}/{6})({d^3\Phi_n}/{dt^3})|_{t=t_{r}}(t-t_{r})^3} \nonumber \\
\fl &\simeq&\left\{
  \begin{array}{@{}ll@{}}
  g(t=t_r) e^{i \Phi_n(t_{r},k_{\|})} R(t_r) & \textrm{if}\  d^3\Phi_n/dt^3 >0 \\
  g(t=t_r) e^{i \Phi_n(t_{r},k_{\|})} R(t_r)^*   & \textrm{otherwise},
    \end{array}\right.     \label{intinairy}
    \end{eqnarray}
    where $t=0$ and $t=T_{p/2}$ denote $t$ at the outer-midplane and inner-midplane, respectively, and the periodicity in a bounce time $T_p$ is assumed for simplicity. Here, $R(t_r)$ is defined by
        \begin{eqnarray}
        R(t_r)=   e^{-i\alpha(x_1t_++t_+^3/3)} Ai\left(x_1,\alpha,t_0\right) \label{Rt},
  \end{eqnarray}
   where the incomplete Airy function is $Ai(x,\alpha,t_0)=\int_{t_0}^{\infty} dt e^{i\alpha(xt+t^3/3)}$, and
          \begin{eqnarray}
     \alpha&=&\left|\frac{1}{2}\frac{d^3\Phi_n}{dt^3} \right|_{t=t_r} ,\;\;
      t_+=\left | \frac{d^2\Phi_n/dt^2}{d^3\Phi_n/dt^3} \right|_{t=t_r}, \nonumber\\    
     x_1&=&-t_+^2 , \;\;\;\textrm{and}\;\;\;\
     t_0=t_+-t_r \label{defvar2}.
       \end{eqnarray}
For the resonance around the outer-midplane (i.e. $t_r \sim 0$), the incomplete Airy function can be approximated by the complete Airy function, which is used in \cite{Bernstein:PF1981,Kerbel:PF1985,Becoulet:PFB1991,Catto:PF1992,Lamalle:PPCF1997,Eester:PPCF2005,Johnson:NF2006}. Around the outer-midplane $t \sim0$, the gyrofrequency $\Omega$ is approximately parabolic in $t$ and it dominates the phase, giving $d\Phi_n/dt \propto (t^2- t_r^2)$. Then the Taylor series up to third order is sufficient, and the resonance happens in the saddle point $t_+=t_r$. In this case, $t_0=0$ and 
  \begin{eqnarray}
Ai\left(x_1,\alpha,t_0=0\right)= \frac{\pi}{\alpha^{1/3}} Ai\left( \alpha^{3/2}x_1\right), 
   \end{eqnarray}
 where the complete Airy function is defined by $Ai(x)=(1/\pi)\int_0^{\infty} dt e^{i(xt+t^3/3)}$\footnotemark[1]. 
       \footnotetext{The real part of the function $Ai(x)$ is determined by the first kind conventional airy function $Ai(x)=\int_0^{\infty} dt \cos(xt+t^3)$, and the imaginary part is determined by $\int_0^{\infty} dt \sin(xt+t^3)=Bi(x)/3+\int_0^{x}(Ai(x)Bi(t)-Bi(x)Ai(t))dt$ where $Bi$ is the conventional second kind airy function.} 
  
  The complete Airy function is useful because it can be applicable to the resonance around the inner-midplane ($t_r\sim T_{p/2}$). In this case, since $d\Phi_n/dt \propto -((t-T_{p/2})^2- t_r^2)$, Eq. (\ref{intinairy}) holds with $t_0=0$ by replacing $t$ variables with $T_{p/2}-t$, and the sign of $\Phi_n/dt$ is opposite to the case of $t_r\sim0$. Thus, we will use the complete Airy function for $R(t_r)$ in the evaluations from Section 3.2.

  For the negative parallel velocity particles in the lower-midplane, the trajectory integral has the same value as Eq. (\ref{intinairy}). On the other hand, for the positive parallel particles in the lower mid-plane or the negative parallel particles in the uppwer-midplane, the integral needs to be be conjugated, 
          \begin{eqnarray}
\fl  && \int_{-T_{p/2}}^{0} dt g(t) e^{i\Phi_n(t,k_{\|})}
\simeq\left\{
  \begin{array}{@{}ll@{}}
  g(t=t_r) e^{i \Phi_n(t_{r},k_{\|})} R(t_r)^* & \textrm{if}\  d^3\Phi_n/dt^3 >0 \\
  g(t=t_r) e^{i \Phi_n(t_{r},k_{\|})} R(t_r)   & \textrm{otherwise}.
    \end{array}\right.      \label{intinairy2}
    \end{eqnarray}
 The conditions to use $R(t_r)$ or $R(t_r)^*$ are summarized in Table 1.

When the resonance point $t_r$ is far from both the outer-midplane and the inner-midplane, the second order term is likely much larger than the third order term in the expansion of $\Phi_n$. It results in the ordinary stationary approximation \cite{Stix:AIP1992} using the second order term,
   \begin{eqnarray}
 \Phi_n (t,k_{\|})&\simeq&  \Phi_n(t_{r},k_{\|})+\frac{1}{2}\frac{d^2 \Phi_n}{dt^2}\bigg|_{t=t_{r}}(t-t_{r})^2, \label{phasebigt}\\
    \int_{0}^{T_{p/2}} dt g(t) e^{i \Phi_n(t,m)}&\simeq& g(t=t_{r})  e^{i \Phi_n(t_{r},k_{\|})}\sqrt{\frac{2\pi}{id^2 \Phi_n/dt^2}}\bigg|_{t=t_{r}} \label{inttbig}.
    \end{eqnarray} 
  This can be also described by the Airy function in the asymptotic limit of $|\alpha^{3/2}x_1|\gg 1$ \cite{Levey:RS1969,Cwik:RS1988},
              \begin{eqnarray}
      && \fl e^{-i\alpha(x_1t_++t_+^3/3)}Ai\left(x_1,\alpha,t_0\simeq 0\right) \simeq \left(\frac{\pi}{\alpha}\right)^{1/2}\frac{1}{(-x_1)^{1/4}} e^{i\pi/4} \;\;\textrm{as}  \;\;\ \alpha^{3/2}x_1 \rightarrow -\infty \nonumber \\
            &=& \left(\frac{2\pi}{id^2\Phi_n/dt^2}\right)^{1/2} \label{inttbiglimit}.
    \end{eqnarray}
    
    The non-resonant interaction can be important only in several specific locations of a flux surface ($t\sim t_{nr}$), where it satisfies $d^2\Phi_n/dt^2\simeq 0$. In a toroidal geometry, for passing particles, $d^2\Phi_n/dt^2\simeq 0$ at outer-midplane ($t_{nr}=0$) and inner-midplane ($t_{nr}=T_{p/2}$) of a flux surface because of the background magnetic field. For trapped particles, it happens at the outer-midplane ($t_{nr}=0$) and two particle tips ($t_{nr}=T_{t1}$ and $t_{nr}=T_{t2}$) where the parallel velocity is zero in the upper-midplane and lower-midplane, respectively. 
    In these locations, even for the non-resonant particles $d\Phi_n/dt\neq0$, the contribution of the phase integral to the diffusion is possibly significant. For the small but non-zero $d\Phi_n/dt$, the correlation length of the plasma and wave interactions is comparable to the length scale of the magnetic fields variation. For example, the expansion around the outer-midplane ($t_{nr}=0$) is 
      \begin{eqnarray}
   \Phi_n(t,k_{\|})\simeq \frac{d \Phi_n}{dt}\bigg|_{t=0}t+\frac{1}{6}\frac{d^3 \Phi_n}{dt^3}\bigg|_{t=0}t^3, \label{phaset0}
  \end{eqnarray} 
  where ${d^2 \Phi_n}/{dt^2}$ is likely to be zero because of the even parity of the magnetic field at the midplane. The integral of the phase in Eq. (\ref{phaset0}) can be evaluated analytically by the complete Airy function, giving 
   \begin{eqnarray}
   \int_{-T_{p/2}}^{T_{p/2}} dt g(t) e^{i \Phi_n(t,m)}&\simeq& g(t_{nr}) e^{i \Phi_n(t_{nr},m)}N(t_{nr}).  \label{intt0}
    \end{eqnarray}
    The non-resonant interaction in $N(t_{nr})$ is evaluated as 
       \begin{eqnarray}
N(t_{nr}) =  \frac{\pi}{\alpha^{1/3}} \left(Ai\left(\alpha^{2/3}x_2\right)+Ai\left(\alpha^{2/3}x_2\right)^*\right),  
  \label{Nt}
 \end{eqnarray}
    where the contribution in the lower-midplane is the conjugate of that in the upper-midplane. Here the new variable $x_2$ is defined by the derivatives at $t=t_{nr}$,
     \begin{eqnarray}
     x_2&=&{2}\left |\frac{d\Phi_n/dt}{d^3\Phi_n/dt^3}\right|_{t=t_{nr}}. \label{defvar1}
       \end{eqnarray}
       Because the Airy function decays to zero at a fast rate as the positive argument increases as shown in Figure 3-(a), the contribution is not negligible only for $\alpha^{2/3}x_2 <1$ and this argument determines the width of the non-resonant contribution in $\mathbf{v}$ space or in $\mathbf{k}$ space from the boundary of the resonant contribution, as will be shown in Sec. 4.2. Around the boundary, the resonant contribution in Eq. (\ref{intinairy}) and the non-resonant contribution in Eq. (\ref{intt0}) are smoothly connected because the both arguments $x_1$ in Eq. (\ref{intinairy}) and $x_2$ in Eq.(\ref{intt0}) converge to zero and the phases other than Airy functions are smoothly continuous, as will be shown in Figure 4-(b). 

  \Table{\label{tab:pinch_diff} Condition for $R(t_r)$ or $R(t_r)^*$ for $ \mathbf{T_n}^*$ in Eq. (\ref{Dql5}) }
\br
&\centre{2}{$t_r$ in upper mid-plane\;\;\;}&\centre{2}{$t_r$ in lower mid-plane} \\
& $v_\|>0$ & $v_\|<0$  &$v_\|>0$& $v_\|<0$\\
\hline
$d^3\Phi_n/dt^3 >0$&  $R(t_r)$ & $R(t_r)^*$& $R(t_r)$ & $R(t_r)^*$\\ \hline
  $d^3\Phi_n/dt^3 <0$&  $R(t_r)^*$ & $R(t_r)$ & $R(t_r)^*$&$R(t_r)$  \\
\br
\end{tabular}
\end{indented}
\end{table}

    \begin{figure} 
    \includegraphics[scale=0.37]{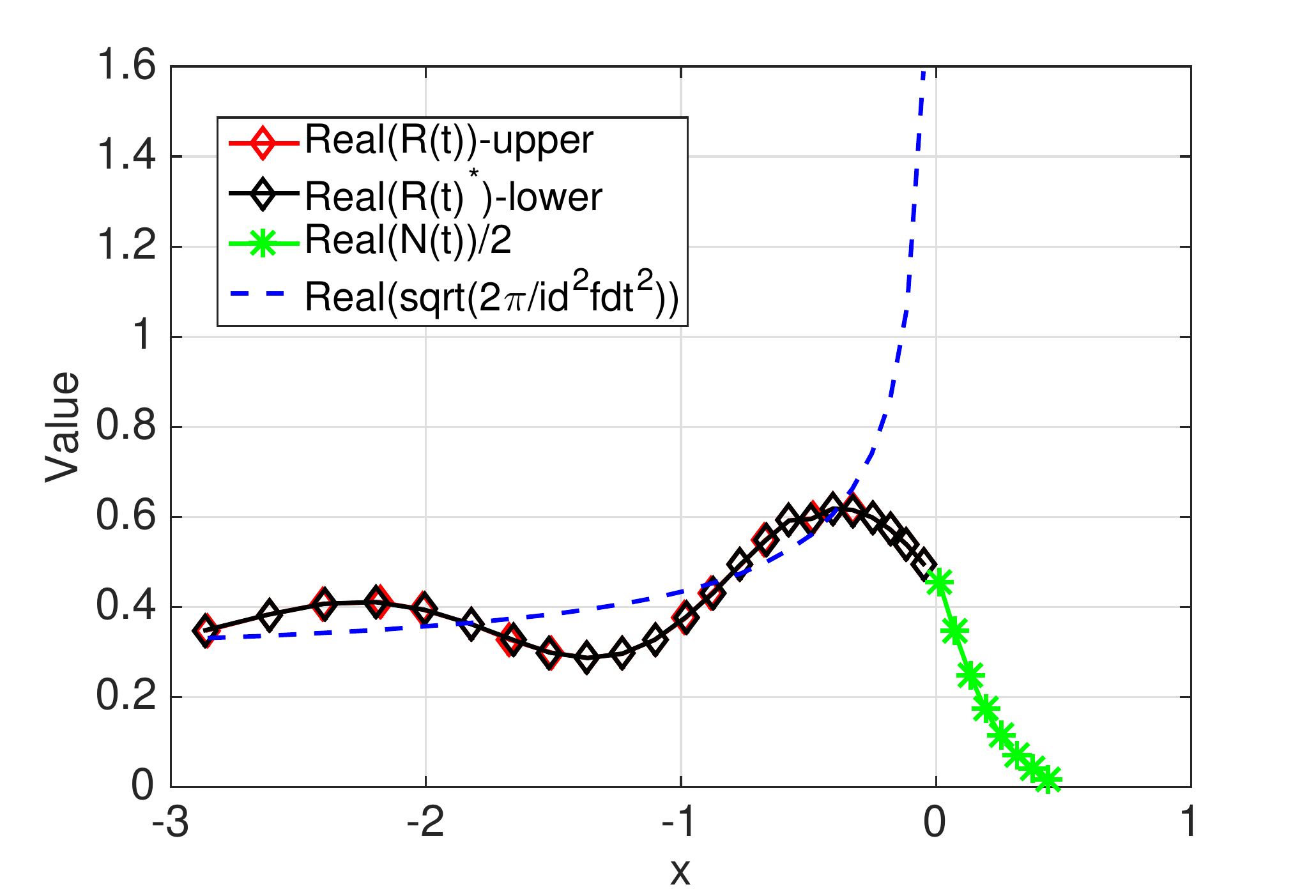}\includegraphics[scale=0.35]{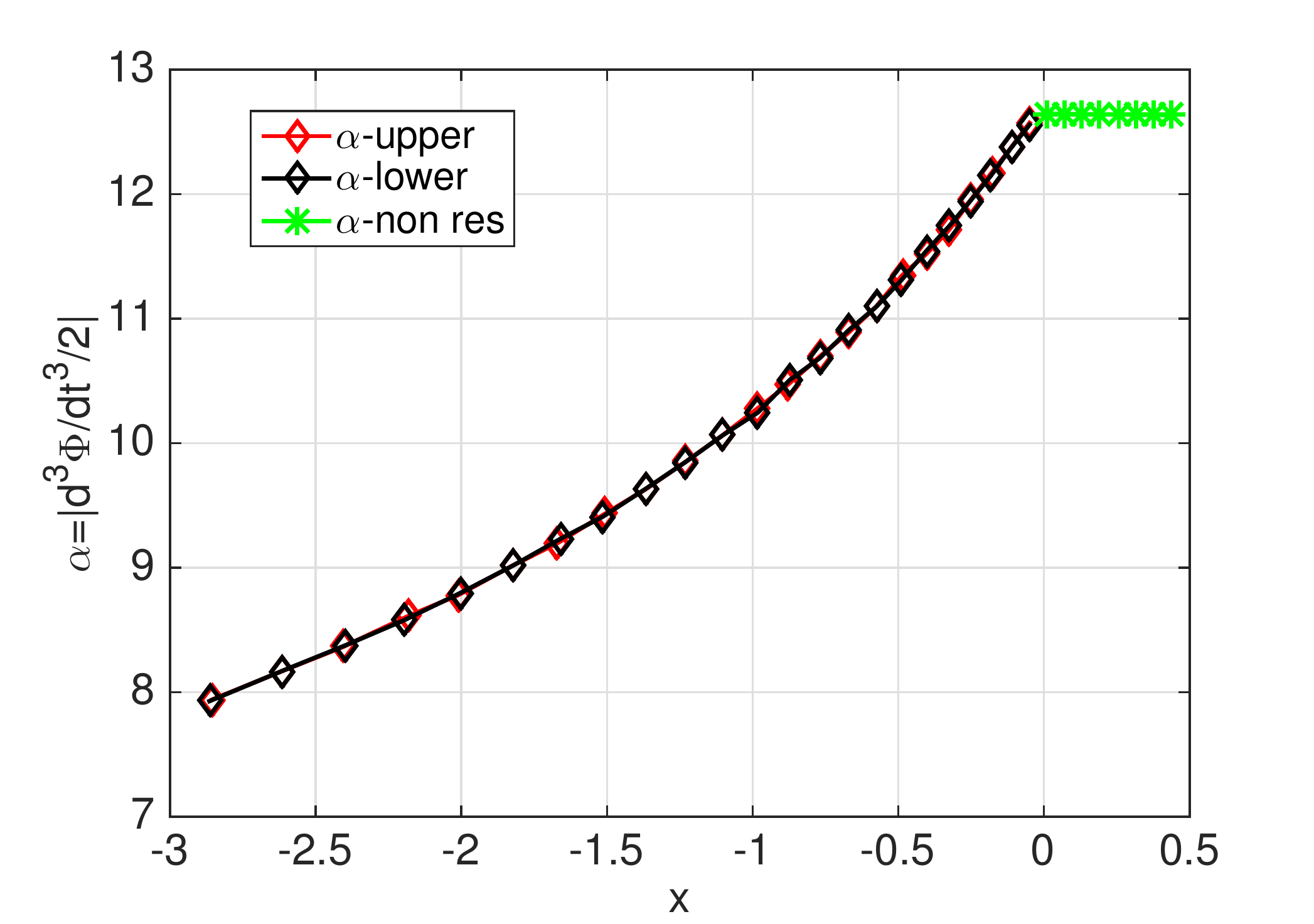}
\caption{ (a) Real part of the trajectory integral in Eq. (\ref{Rt}) and Eq. (\ref{Nt}) in terms of the parameter x of adjacent poloidal modes, and (b) the variation of parameter $\alpha$ in terms of x of adjacent poloidal modes. Each data point is obtained for the consecutive spectral modes numbers with a fixed velocity variable $(E,\mu)$ in a circular tokamak. If the spectral mode is in the group $\mathcal{K}_{r}$, $x=x_1$ and Figure (a) shows $R(t)$ in Eq. (\ref{Rt}) in the upper mid-plane (red) and in the lower mid-plane (black) with its asymptotic limit in Eq. (\ref{inttbiglimit}) (blue). If the spectral mode is in the group $\mathcal{K}_{nr}$, $x=x_2$ and Figure (a) shows $N(t)$ in Eq. (\ref{Nt}) (green). 
}
\label{fig:d_opt}
\end{figure}

    \begin{figure} 
    \includegraphics[scale=0.34]{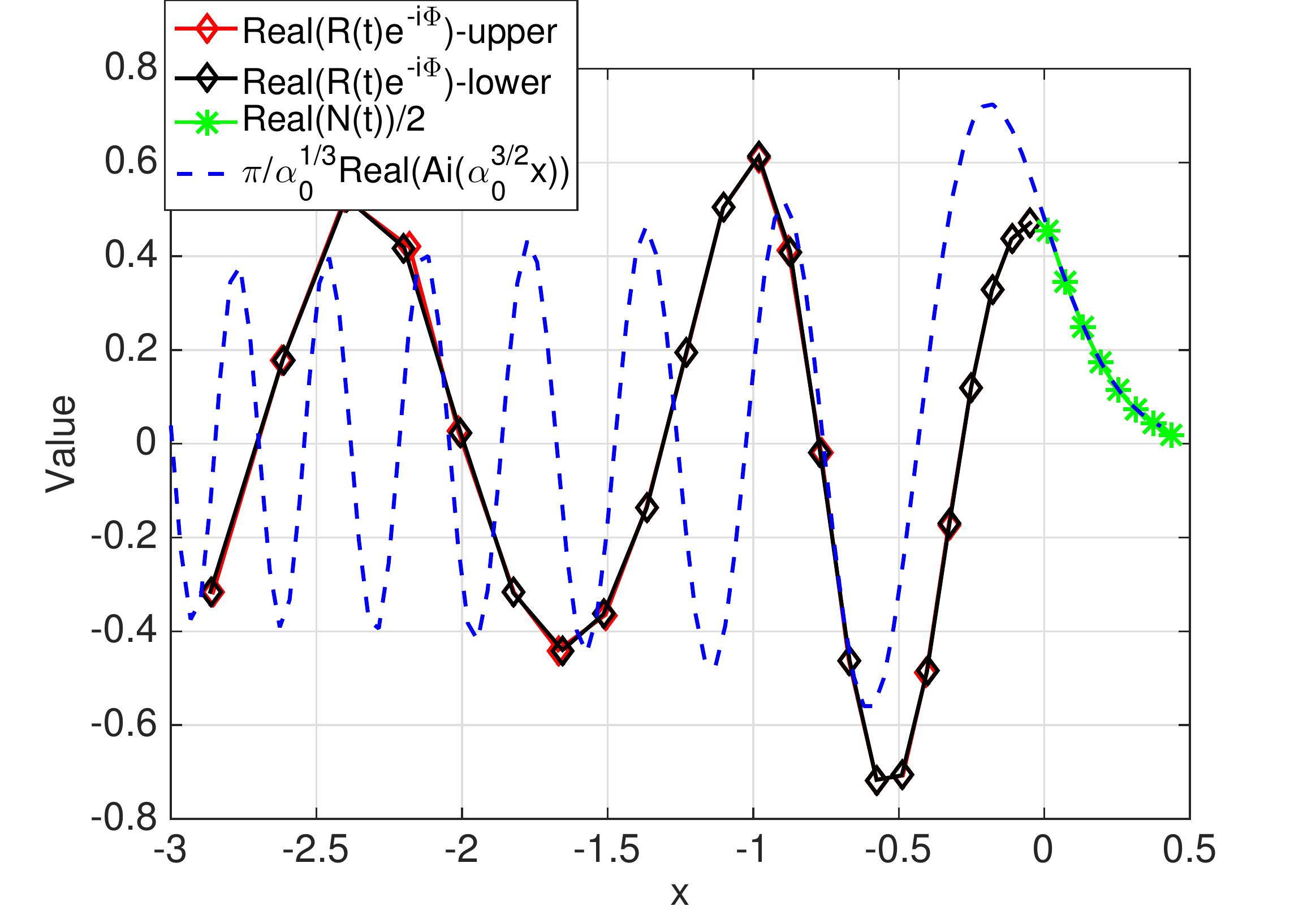}\includegraphics[scale=0.37]{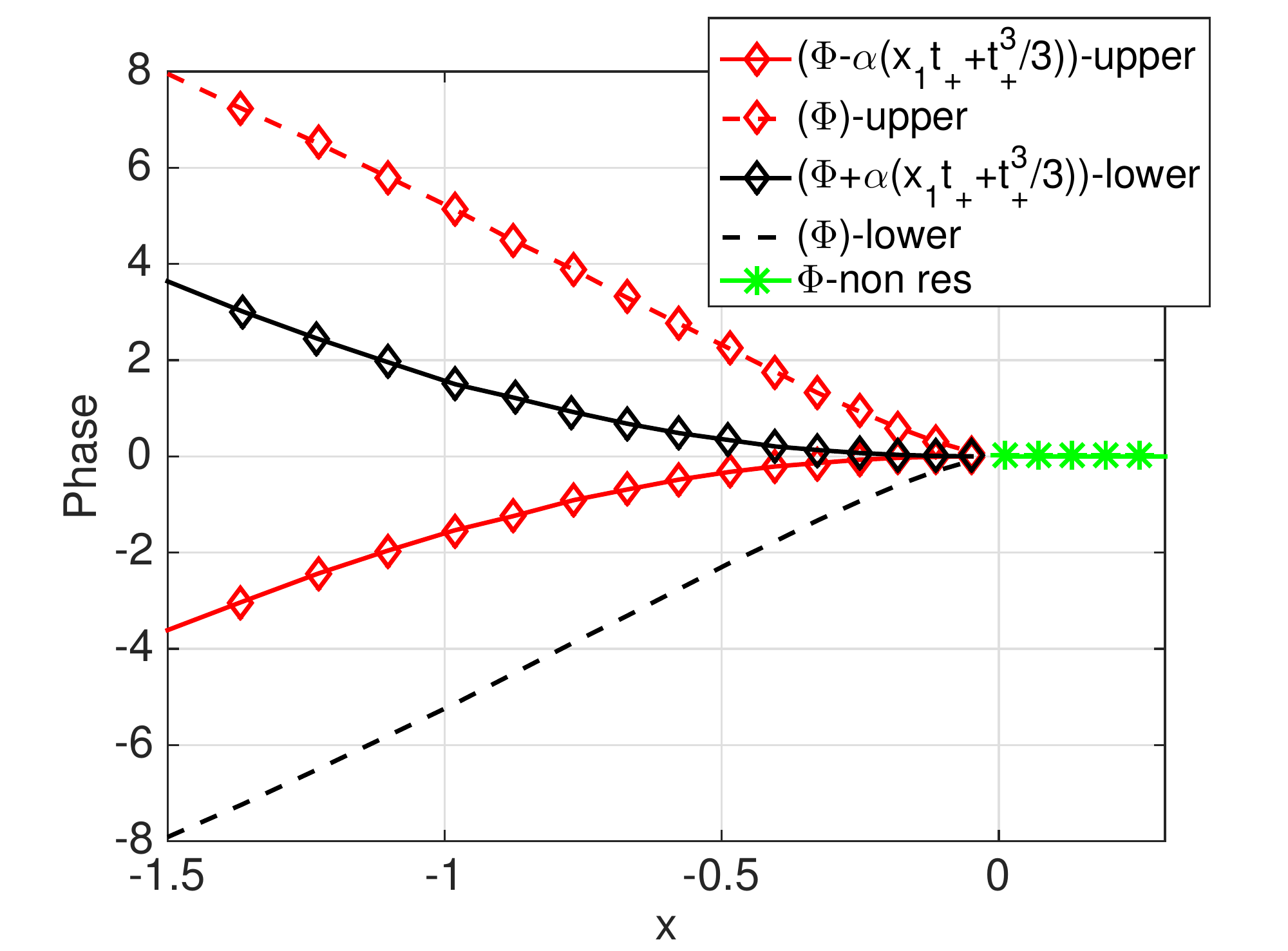}
\caption{ (a) Real part of $R(t_r)\exp(i\Phi_n(t_r))$ and $N(t_{nr})\exp(i\Phi_n(t_{nr}))$ and (b) the phase $\Phi_n$ in terms of the parameter x of adjacent poloidal modes for the same example as Figure 3. Here, $t_{nr}=0$ and $\Phi_n(t_{nr})=0$ in the outer-midplane. 
}
\label{fig:d_opt}
\end{figure}

  \subsection{Implementation}\label{sec3.2}
  
  In this section, we explain how to implement the positive-definite form of the bounce averaged quasilinear diffusion coefficients in the wave code TORIC \cite{Brambilla:PPCF1999}.  Before explaining the implementation, we comment about the conductivity tensor, which has the general relation of Eq. (\ref{rel1}) with the diffusion coefficients. Since both the conductivity tensor used in the Maxwell's equation and the diffusion coefficients used in the Fokker-Planck equation are based on the derivation of the perturbed distribution by the RF waves (e.g. Eq. ($\ref{f_k2}$)), they need to be consistent in terms of the assumption of the derivations. For example, in our previous paper \cite{Lee:PoP2017}, we have derived the quasilinear diffusion in the homogenous plasmas and magnetic fields in a small Larmor radius approximation, which is equivalent to the assumption of the dielectric tensor used in TORIC \cite{Brambilla:PPCF1999}. The consistent conductivity tensor and quasilinear diffusion in Eq. (23) and (34) of \cite{Lee:PoP2017} guarantee the same power absorption $ \langle \mathbf{J}\cdot \mathbf{E} \rangle_w =\dot{W}$ in the lowest order of the small Larmor radius expansion \cite{Lee:PoP2017}, which is useful for the convergence of the iteration between TORIC and CQL3D. However, in this paper, we modified the existing quasilinear diffusion in \cite{Lee:PoP2017} to include the parallel inhomogeneity, while keeping the existing dielectric tensor in TORIC. Because the trajectory integral with the Airy function is not consistent with the plasma dispersion function of which imaginary part corresponds to the Dirac-delta function, we have to redefine the dielectric tensors. We remain it as a future work since it is not trivial to find the pre-calculated velocity integral values for the conductivity without the Dirac-delta function, and it is computationally expensive. There were some previous work to modify the plasma dispersion function to consider the parallel inhomogeneity \cite{Brambilla:PLA1994, smithe:PRL1988, Berry:PoP2016}, but their dielectric tensor are not exactly consistent with the quasilinear diffusion in this paper.

  Using the approximation of the trajectory integral in Section 3.1, the positive-definite form of the bounce-averaged quasilinear diffusion coefficient can be defined by
     \begin{eqnarray}
&&\mathbf {T_n}^*(E, \mu)=    \int_{\mathbf{k}\in \mathcal{K}_r} d\mathbf{k} \sum_{r} \mathbf{G}( \mathbf{P_n}(t_{r}) \cdot \mathbf{E(\mathbf{k})})^*e^{i \Phi_n(t_{r},k_{\|})} R(t_r)\nonumber \\
&& +\int_{\mathbf{k}\in \mathcal{K}_{nr}} d\mathbf{k}\sum_{nr} \mathbf{G}(\mathbf{P_n}((t_{nr}) \cdot \mathbf{E(\mathbf{k})})^*e^{i \Phi_n(t_{nr},k_{\|})}N(t_{nr}), 
\label{Dql5}
 \end{eqnarray}
 where $\mathcal{K}_{r}$ is the group of the spectral modes that have at least one location of a flux surface satisfying the resonance condition and $\sum_{r}$ is the summation of the values at all resonant positions $t_r$, while $\mathcal{K}_{nr}$ is the group of the spectral modes that do not satisfy the resonance condition in all locations of the flux surface and $\sum_{nr}$ is the summation of the values at the specific positions $t_{nr}$ for the non-resonant interactions. As mentioned in the previous section, the location of $t_{nr}$ is determined by the possible positions where the correlation length of the plasma-wave interaction is comparable to the length for the magnetic fields variation. For passing particles, $t_{nr}$ is located in the outer-midplane ($t_{nr}=0$) and the inner-midplane ($t_{nr}=T_{p/2}$), and for trapped particles, $t_{nr}$ is located in the outer-midplane ($t_{nr}=0$) and two tips of the trapping in the upper-midplane and the lower-midplane. Here, the resonant interaction in $R(t)$ is replaced by its conjugate according to the conditions in Table 1.

 To be coupled with CQL3D, we reformulate the quasilinear diffusion tensor in a spherical coordinate, $(v, \vartheta, \phi)$, where $v=\sqrt{v_\perp^2+v_\|^2}$ is the speed, $\vartheta=\arctan(v_\perp/v_\|)$ is the pitchangle, and $\phi$ is the gyroangle. Then, the quasilinear diffusion coefficients ($\textsf{B}$, $\textsf{C}$, $\textsf{E}$ and $\textsf{F}$) determine the divergence of the flux in velocity space by \cite{Kerbel:PF1985}. After taking a bounce average of the gyroaveraged quasilinear diffusion using the invariant variables defined at the outer-midplane $(v, \vartheta_0)$, the bounce-averaged diffusion for CQL3D is
 \begin{eqnarray}
 \fl &&\lambda_p \langle Q(f) \rangle_b =\frac{1}{v^2} \frac{\partial }{\partial v}\left \lbrace \lambda_p {\langle \textsf{B}} \rangle_b \frac{\partial }{\partial v}+\lambda_p \left\langle \textsf{C}\frac{\partial \vartheta_0 }{\partial \vartheta} \right\rangle_b \frac{\partial }{\partial \vartheta_0} \right \rbrace f (v, \vartheta_0)  \nonumber\\ 
\fl && +\frac{1}{v^2 \sin \vartheta_0} \frac{\partial }{\partial \vartheta_0}\left \lbrace \lambda_p \left \langle {\textsf{E}}\frac{\sin \vartheta_0}{\sin \vartheta}\frac{\partial \vartheta_0 }{\partial \vartheta}  \right\rangle_b \frac{\partial }{\partial v}+ \lambda_p \left \langle {\textsf{F}}\frac{\sin \vartheta_0}{\sin \vartheta}\left(\frac{\partial \vartheta_0 }{\partial \vartheta} \right)^2 \right\rangle_b \frac{\partial }{\partial \vartheta} \right \rbrace f (v, \vartheta_0).
 \end{eqnarray}

 TORIC uses the dielectric tensor in the expansion by a small parameter of $k_\perp \rho_i$ \cite{Brambilla:PPCF1999}, and in our previous paper \cite{Lee:PoP2017} we derived the bounce-averaged diffusion coefficient to the lowest order for the fundamental cyclotron damping ($n=1$) in Eq. (44) of  \cite{Lee:PoP2017} as 
 \begin{eqnarray}
 \langle \textsf{B}\rangle_b &=& \frac{\pi \epsilon \omega_p^2}{ 4 m n_s}\frac{1}{T_p} \int_0^{T_p} dt \sum_{m_1} \sum_{m_2} Re \bigg[e^{i(m_2-m_1)\theta} {E_+}(m_1) \nonumber\\
&&\times \frac{v_\perp^2 c^2}{|k_\||}\delta\left(v_\|-\frac{\omega-\Omega}{k_\|} \right) {E_+}(m_2)\bigg] \label{lambdaB1},
\end{eqnarray}
 where the electric field is decomposed in TORIC into poloidal spectral modes $\sum_{m} \mathbf{E}(m)\exp({im\theta})$ for a fixed toroidal spectral mode at each radial element. Here, the phase is determined by $\omega-n\Omega-k_\|v_\|$ where the poloidal mode $m$ is used to determine $k_\|v_\|= m\dot{\theta}+n_\varphi\dot{\varphi}$, while the Kaufman form uses  $\omega-n\Omega-j\omega_b-n_\varphi\dot{\varphi}$, where $n_\varphi$ is the toroidal mode number, $\dot{\theta}$ and $\dot{\varphi}$ are the poloidal and toroidal velocity rate, respectively and $\omega_b$ is the bounce frequency. In TORIC, the toroidal spectral modes are decoupled each other for the toroidal axisymmetric geometry, while the poloidal spectral modes are coupled by the weak form of the wave equation on the finite element code for the inhomogeneous dielectric tensor and parallel wave vector $k_\|(\theta)$ along the poloidal angle \cite{Brambilla:PPCF1999}.  Eq. (\ref{lambdaB1}) is used to produce the results for the K-E diffusion coefficient in Figure 1. 
 
In this paper, we modify the coefficient $\langle \textsf{B} \rangle_b$ for the positive-definite form, giving
  \begin{eqnarray}
 \langle \textsf{B}\rangle_b &=& \frac{\epsilon c^2 \omega_p^2}{ 8 m n_s}\frac{1}{T_p}\bigg[   \sum_{m \in \textsf{m}_r}  \sum_{r}  {E_+}^*(m) v_\perp(t_r) e^{i \Phi_n(t_{r},k_{\|})} R(t_r)\nonumber \\
&& + \sum_{m \in \textsf{m}_{nr}}  \sum_{nr} {E_+}^*(m) v_\perp(t_{nr})e^{i \Phi_n(t_{nr},k_{\|})}N(t_{nr}) \bigg]^2, \label{DqlB}
\end{eqnarray}
where $\textsf{m}_r$ is the group of the poloidal modes that have at least one location of a flux surface satisfying the resonance condition, while $\textsf{m}_{nr}$ is the group of the poloidal modes that do not satisfy the resonance condition in all locations of the flux surface. Because we use the same diffusion vector $\mathbf{G}$ as the Kennel-Engelmann diffusion, the relations between $\langle \textsf{B} \rangle_b$ with other coefficients by $\textsf{C}$, $\textsf{E}$, and $\textsf{F}$ are the same as Eq. (50) of \cite{Lee:PoP2017}. Also, as mentioned, we keep the dielectric tensor as in \cite{Lee:PoP2017,Bertelli:NF2017} (e.g. $(\hat{L}+ \Delta\hat{L}_1)E_+$ in Eq. (25) and (26) of \cite{Lee:PoP2017} for the fundamental cyclotron damping).

Figure 3-(a) shows the evaluation of $R(t_r)$ in Eq. (\ref{Rt}) and $N(t_{nr})$ in Eq. (\ref{Nt}) for different poloidal mode numbers of TORIC. In the figure, if the poloidal mode number has at least one resonance (i.e. $m \in \textsf{m}_r$), $R(t_r)$ is shown in $x=x_1$ at $t=t_r$, and otherwise (i.e. $m \in \textsf{m}_{nr}$) $N(t_r)$ is shown in $x=x_2$ at $t=t_{nr}$. For the evaluation of Figure 3, we use the up-down symmetric flux surface that has the cyclotron resonance layer close to the the outer-midplane $t=0$. Thus, the range of the evaluation is selected as [-$T_{p/2}$,$T_{p/2}$] with $t_{nr}=0$. The real part of $R(t_r)$ in the upper mid-plane is the same as that in lower mid-plane because of the up-down symmetry, and the imaginary parts of $R(t_r)$ cancel each other. The real part of $R(t_r)$ is smoothly connected with $N(t_{nr})/2$ around $x=0$ due to their definitions using Airy functions. However, note that $R(t_r)$ has the additional factor, $\exp({-i\alpha(x_1t+t^3/3)})$, other than the Airy function, and because of the factor it converges in the limit of the stationary approximation asymptotically when $\alpha^{3/2}x_1 \rightarrow -\infty$, as shown in Eq. (\ref{inttbiglimit}) (see the blue graph of Fig 3-(a)). Figure 3-(b) shows the variation of $d^3\Phi_n/dt^3$ at $t=t_r$ or $t=t_{nr}$ due to the change of the magnetic field in $t$.

Figure 4-(a) shows the values of $R(t_r)\exp(i\Phi_n(t_r))$ that has the additional phase factor from the values in Figure 3-(a). The phase $\Phi_n(t_r)$ determines the correlation length of the contributions. The fast oscillation by the phase change in $\Phi_n(t_r)$ as the resonance location becomes away from $t=t_{nr}=0$ results in the decorrelation between each resonance. As shown in Figure 4-(a), when $t_r$ is away from the $t=0$, each resonance is likely decorrelated from each other due to the phase mixing \cite{Stix:AIP1992}. However, around $t=0$ the phase mixing is significantly reduced and each resonance is likely correlated because there is a substantial cancellation of the phase between $\Phi_n$ and $\alpha(x_1t_++t_+^3/3)$ of $R(t_r)$. As shown in Figure 4-(b), around $t=0$, the parabolic change of $d\Phi_n/dt \propto (t^2- t_r^2)$ results in the exact cancellation,
     \begin{eqnarray}
\Phi_n(t_{r},k_{\|})+\alpha(x_1t_++t_+^3/3)=0 \;\;\textrm{as}  \;\;\ t_r \rightarrow 0. \label{cancel1}
   \end{eqnarray}
 Thus, only higher order variations (e.g. $t^4- t_r^4$) can contribute the finite phase of $\Phi_n(t_{r},k_{\|})+\alpha(x_1t_++t_+^3/3)$ around $t=0$ for the up-down symmetric geometry. The reduced phase mixing due to the cancellation is an important effect of the toroidal geometry that is captured in the form of this paper but not in the Dirac-delta function of the K-E coefficients that determines the correlation only by the phase $\Phi_n(t_{r},k_{\|})$. In other words, the positive definite form of this paper has the minimized phase mixing of the solid lines in Figure 4-(b), while K-E coefficients have the overestimated phase mixing by the dashed line of Figure 4-(b). Additionally, this cancellation results in the smooth connection between $R(t_r)\exp(i\Phi_n(t_r))$ and  $N(t_{nr})$, and the two consecutive resonances around $t=0$ are likely correlated. In Section 4-1, we will show how the correlation around $t=0$ is important in the evaluation of the diffusion coefficients.
  
The assumption in Section 3.1 for the periodicity of $\Phi_n$ in every bounce time $T_p$ is not generally valid in many flux surfaces. The change of $\Phi_n$ in a period is determined by $\Delta \Phi_n=\Phi_n(t=0)-\Phi_n(t=-T_p)\simeq \int_{-T_p}^0 (\omega-n\Omega(t))dt + (m_s+n_sq)2\pi$ where $m_s+n_sq$ is the parallel spectral mode number of the flux surface, $m_s$ and $n_s
$ are the poloidal and toroidal spectral mode numbers, respectively, and $q$ is the safety factor. All terms of $\Delta \Phi_n$ except the poloidal mode contribution by $m$ result in the non-periodicity of $\Phi_n$ (i.e. $mod(\Delta \Phi_n, 2\pi)\neq0 $ where $mod(a,b)$ is the remainder after division a by b). Thus, the evaluation value of Eq. (\ref{Dql5}) is different depending on the initial value of $t$ in the range of the integration, and the average value of the evaluation in many periods is different depending on the number of periods, although the average value is expected to converge in many periods due to any decorrelation. We will show the importance of the initial value and the number of periods in the examples of Section 4.

For the evaluation in one period $T_p$, the initial value and the last value of $t$ determine the discontinuous point of $\Phi_n$. As shown in Figure 3 and 4, the phase of $\Phi_n$ is important to determine the correlation of the evaluation. Thus, it needs to select the discontinuous point carefully for the implementation. For ICRF waves, if the cyclotron resonance layer is located in the major radius larger than the magnetic axis, the correlation around the outer-midplane is more important than that around the inner-midplane. In that case, the evaluation in the range [-$T_{p/2}$,$T_{p/2}$] is desirable to have the continuous $\Phi_n$ around the outer-midplane $t=0$ (or $\theta=0$). In other words, if the phase $\Phi_n$ is calculated in the poloidal angle from $\theta=-\pi$ to $\theta=\pi$ for passing particles (the reference poloidal angle $\theta_{ref}=-\pi$) and from $\theta=2\pi-\theta_{t2}$ to $\theta=\theta_{t1}$ for trapped particles, where $\theta_1$ and $\theta_2$ are the poloidal angles of trapping tips with $0<\theta_1<\theta_2<2\pi$. On the other hand, the major radius of the resonance layer is less than the magnetic axis, the continuous $\Phi_n$ around the inner-midplane is more important and the range [0,$T_{p}$] (i.e. from $\theta=0$ to $\theta=2\pi$) is desirable. In this case, the reference poloidal angle for the $\Phi_n$ calculation is at the outer-midplane $\theta_{ref}=0$. The significant difference in the diffusion coefficients made by different reference poloidal angles will also be shown in Figure 5-(a).

The evaluation of the positive-definite form is computationally more expensive than that of the K-E coefficients. The evaluation of the form in Eq. (\ref{Dql5}) requires $O(n_r n_v^2 n_{m}^3n_{ch} n_{int})$ floating point operations, while the K-E form in Eq. (\ref{Dql2}) requires $O(n_r n_v n_{m}^2)$ operations.
Here $n_r$ and $n_{m}$ are the number of radial coordinate grid and poloidal coordinate grid points (or poloidal spectral modes), respectively, and $n_v$ is the number of the velocity space grid points in each direction. In the K-E form, the Dirac-delta function reduces the $n_v$ operations, and the integral in $\mathbf{k_2}$ can be evaluated separately from the integral in $\mathbf{k_1}$ \cite{Jaeger:NF2006} so there is a reduction by a factor of $n_m$ operations. However, in the positive-definite form, those reductions are not applicable and there are additional $n_{ch} n_{int}$ operations for the phase evaluation for $\Phi_n$, where $n_{ch}$ is the order of the interpolation (e.g. Chebyshev interpolation) and $n_{int}$ is the number of the required interpolations. For $n_m\sim100$, $n_v\sim100$, $n_{ch}\sim10$, and $n_{int}\sim 10$,  the computation of the positive-definite form is expensive being about $10^{6}$ times more than the K-E form. 

To reduce the computation cost for the high resolution case with a large $n_m$, we introduced the reduced poloidal space grid by $n_\theta\ll 2n_m$ for the evaluation of the trajectory integral, while keeping the original number $n_m$ for the poloidal spectral mode summation. It is because the trajectory integral does not require the fine mesh as much as the shortest wavelength does. It results in the reduction of the computation cost by about 100 times, giving the operation count $O(n_r n_v^2 n_{m}n_{\theta}^{2}n_{ch} n_{int})$.

\section{Features of the diffusion}\label{sec:4}
In this section, we investigate some features of the quasilinear diffusion that is derived in Section 3 and implemented in TORIC. As an example, the previous benchmark study of ICRF waves in \cite{Lee:PoP2017,budny2012benchmarking} is examined for the minority species heating scenarios in ITER with a static magnetic field 5.3T at the magnetic axis. In this example, we simulate three ion species with the density ratio of (D,T,He3)$=(48,48,2)\%$ and the ICRF wave frequency is 50MHz. The dominant wave power is absorbed by the minority species He3 in the off-axis because the cyclotron layer is located at $R=R_0+0.55 \textrm{m}$ on the low field side, which is tangential to the flux surface of $r/a=0.32$. Here, $R_0=6.2 \textrm{m}$ is the major radius of magnetic-axis and $r/a$ is the normalized radial coordinate, which is defined by the square root of normalized poloidal flux. 

Using this example, the positive-definite form of quasilinear diffusion in this paper is compared with the K-E diffusion. To focus on the value of the quasilinear diffusion and exclude the effect of non-Maxwellian, we assume that the total power damping is small ($P_{abs}=1W$) and the distribution functions of all plasma species are approximately Maxweliian. TORIC is used to evaluate the postive-definite quasilinear diffusion coefficients in this paper. In the example, the power decomposition by $\langle \mathbf{E}\cdot \mathbf{J} \rangle_w$ is $59\%$ of He3 fundamental damping, $17\%$ of T second harmonic damping, and 24\% electron damping in TORIC.

\subsection{Correlation between resonances}\label{sec:4.1}
 Figure 5-(a) shows the effect of correlation between consecutive resonances on the power absorption profile. The blue curve of Figure 5-(a) is the power profile by $\langle \mathbf{E}\cdot \mathbf{J} \rangle_w$ and it is supposed to be the same theoretically as the green curve by $\dot{W}$ using the K-E coefficients in the lowest order of the small Larmor radius approximation \cite{Lee:PoP2017}. The difference in the two curves is due to the error introduced by negative values in the quasilinear diffusion as shown in Figure 1. 
 The red and cyan curves of Figure 5-(a) show the power absorption by the positive-definite diffusion coefficients of this paper with each curve having a different poloidal reference $\theta_{ref}$ for the evaluation range of the trajectory integral. As shown in the solid lines of Figure 4-(b), the phase mixing around the outer-midplane ($\theta=0$) is so small that two consecutive resonances in a trajectory around the outer-midplane are likely correlated. In the red curve of Figure 5-(a), by locating the discontinuous poloidal reference location of $\Phi_n$ far from the outer-midplane ($\theta_{ref}=-\pi$), the correlation between two resonances (e.g. $\theta=0.1$ and $\theta=-0.1$) are well captured in the diffusion. On the other hand, in the cyan curve of Figure 5-(a), when the discontinuous poloidal reference is on the outer-midplane ($\theta_{ref}=0$), the two resonances around the outer-midplane (e.g. $\theta=0.1$ and $\theta=2\pi-0.1$) are likely uncorrelated. According to the fundamental theory of statistics, the contributions of the perfectly correlated two kicks on the diffusion is twice larger than that of perfectly uncorrelated two random kicks. The reduced diffusion coefficients can explain the decrease in the power absorption of the cyan curve compared to the red curve around $r/a=0.25$, where the flux surface is almost tangential to the resonance layer around the outer-midplane.  In a real experiment, the actual velocity diffusion by two consecutive resonances is likely determined by neither a perfect correlation of the red curve result or a perfect decorrelation of the cyan curve result. Instead, it could be a point between two case results, which is determined by the dominant decorrelation mechanism in the experiment. 
 
It is worth noting that the blue curve in Figure 5-(a) for $\langle \mathbf{E}\cdot \mathbf{J} \rangle_w$ is similar to the red curve for the correlated resonance, although $\langle \mathbf{E}\cdot \mathbf{J} \rangle_w$ is based on the more decorrelated resonances around the outer-midplane. The Dirac-delta function of the K-E form that is equivalent to $\langle \mathbf{E}\cdot \mathbf{J} \rangle_w$ results in a large value at $\theta\simeq 0$ where $d^2\Phi_n/dt^2 \simeq 0$, as shown in the singularity of the blue curve in Figure 3-(a), which may make the similarly large contribution to the diffusion as the correlated resonance of the red curve in Figure 5-(a). However, it is not necessary that the diffusion with the correlated resonances is always similar to that of the K-E form or $\langle \mathbf{E}\cdot \mathbf{J} \rangle_w$ because this toroidal effect is captured only in the positive-definite form.

 Figure 5-(b) shows that the power profiles are the same for the different velocity space grid resolutions unlike Figure 1 by the K-E diffusion coefficients. It is because the positive definite form does not need the average of negative values in a grid spacing.  
 
Figures 6 and 7 show the contour plots for the component of bounce-averaged quasilinear diffusion tensor in the speed direction, $\langle B \rangle_b =\hat{\mathbf{v}} \cdot \langle D_{ql} \rangle_b \cdot \hat{\mathbf{v}}$, at $r/a=0.25$ in linear scale and log scale, respectively, where $\hat{\mathbf{v}}$ is the unit vector of the velocity. The size and the patterns of all subplots in Figure 6 and 7 are similar but have some distinctive features depending on the evaluation method. The K-E diffusion in Figures 6-(a) and 7-(a) show the discontinuous and noisy patterns (e.g.  discontinuous white parts in the log plot) because of the negative values in the diffusion coefficients, as explained Section 2. Figures 6-(b) and 7-(b) for the positive-definite form with $\theta_{ref}=-\pi$ show more smooth and distinctive contour patterns due to the correlated resonances than those of Figures 6-(c) and 7-(c) for the positive-definite form with $\theta_{ref}=0$.

 \begin{figure*}
(a) \;\;\;\;\;\;\;\;\;\;\;\;\;\;\;\;\;\;\;\;\;\;\;\;\;\;\;\;\;\;\;\;\;\;\;\;\;\;\;\;\;\;\;\;\;\;\;\;\;\;\;\;\;\;(b) \;\;\;\;\;\;\;\;\;\;\;\;\;\;\;\;\;\;\;\;\;\;\;\;\;\;\;\;\;\;\;\;\;\;\;\;\;\;\;\;\;\;\;\;\;\;\;\;\;\;\;\;\;\;\;\;\;\;\;\;\;\;\;\\
\includegraphics[scale=0.4]{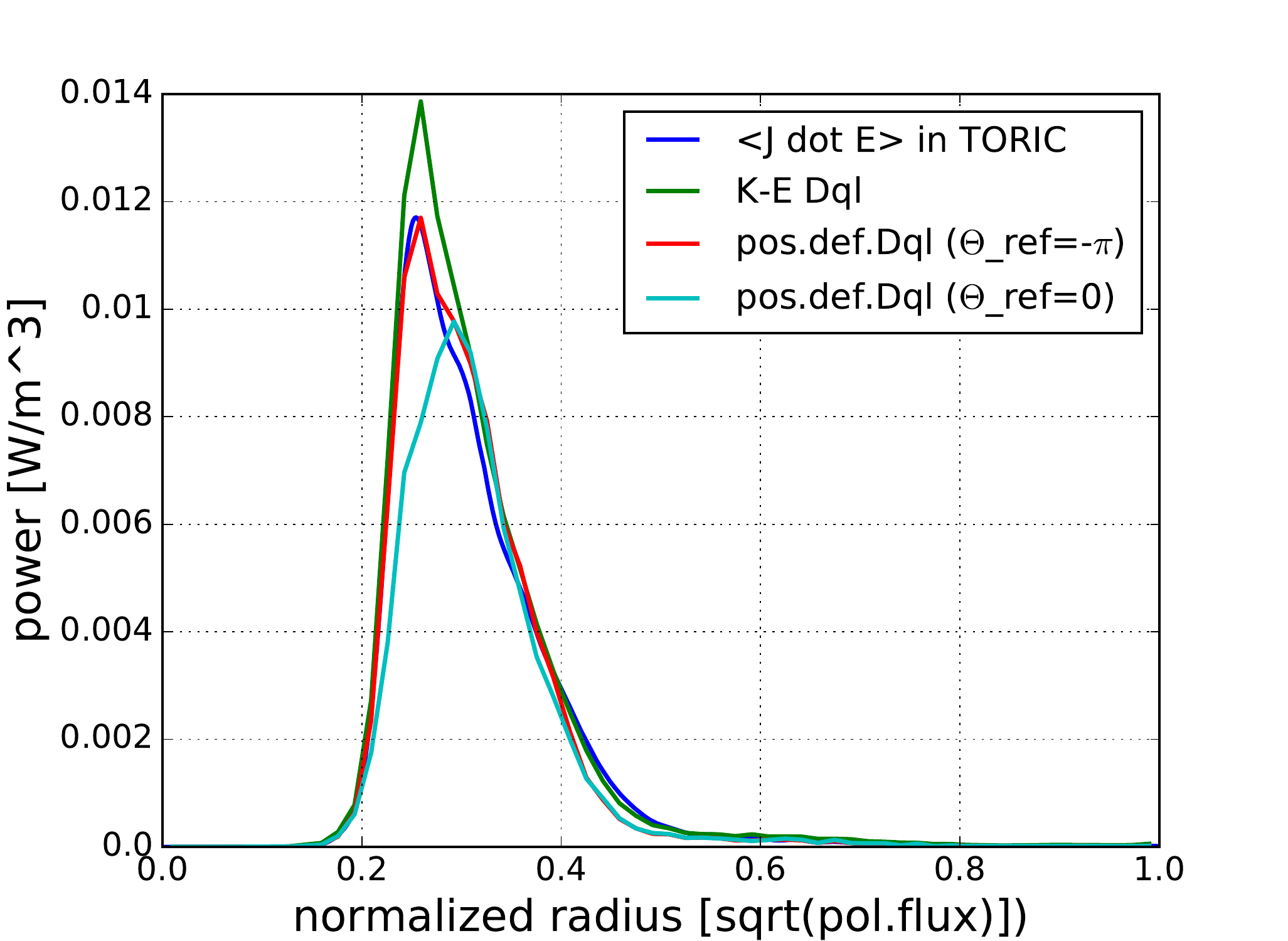}\includegraphics[scale=0.4]{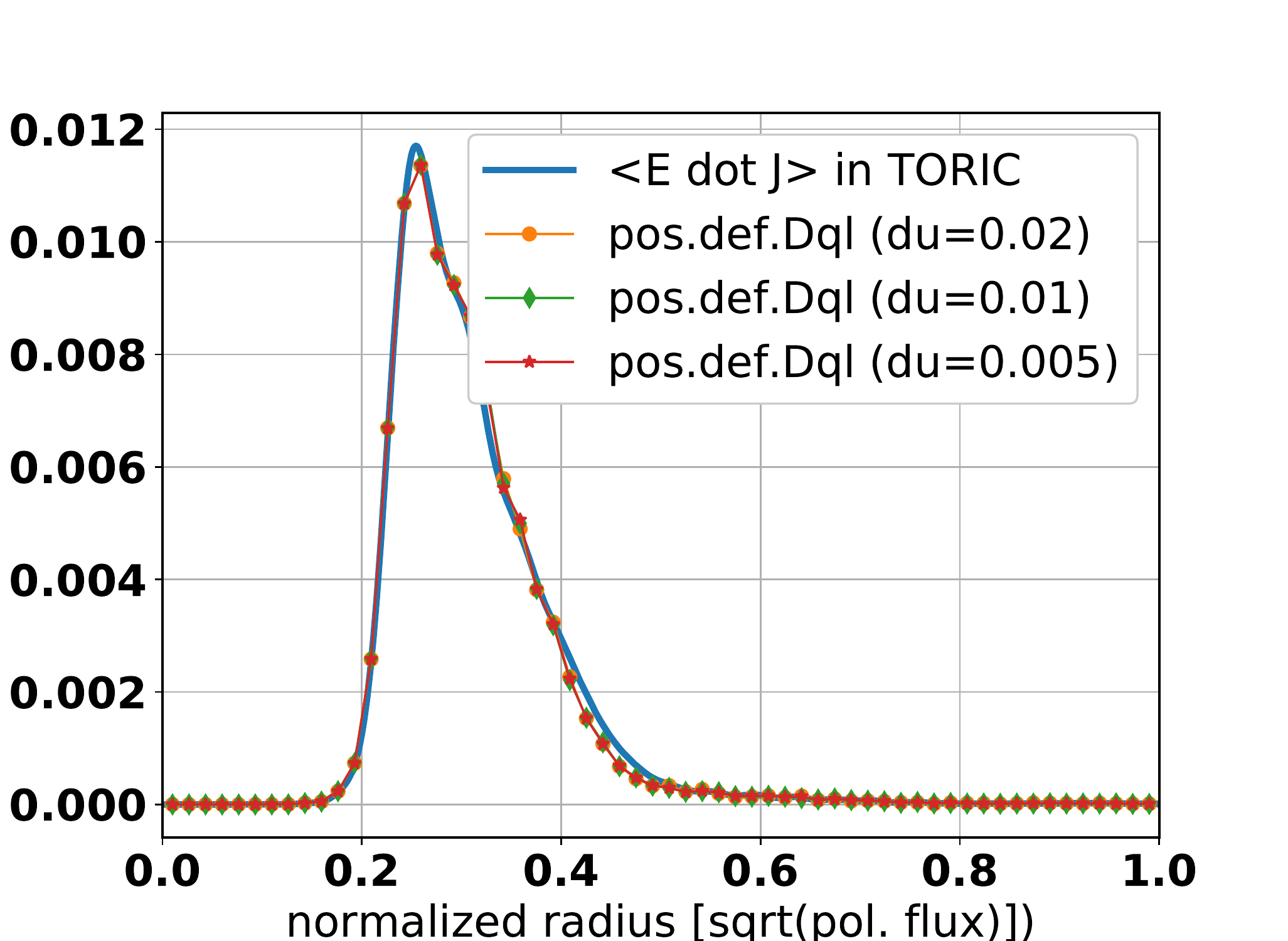}
\caption{Radial profiles of power absorption by He3 for the 50 MHz ICRF injection in ITER. The profiles are simulated by $ \langle \mathbf{E}\cdot \mathbf{J} \rangle_w$ (blue) in TORIC and $\dot{W}$ with different quasilinear diffusion coefficients (other colors). In (a), the green curve uses the Kennel-Engelmann coefficients, and the red and cyan curves use the positive-definite coefficients with $\theta_{ref}=-\pi$ and $\theta_{ref}=0$, respectively. The velocity space grid resolution for all diffusion coefficients is  $\Delta v=0.01$.  In (b), the profiles are obtained by the positive-definite coefficients with $\theta_{ref}=-\pi$ and various velocity space resolutions for $\Delta v=0.02$,  $\Delta v=0.01$,  and $\Delta v=0.005$.} 
\end{figure*}

 \begin{figure*}
(a)\\
\includegraphics[scale=0.2]{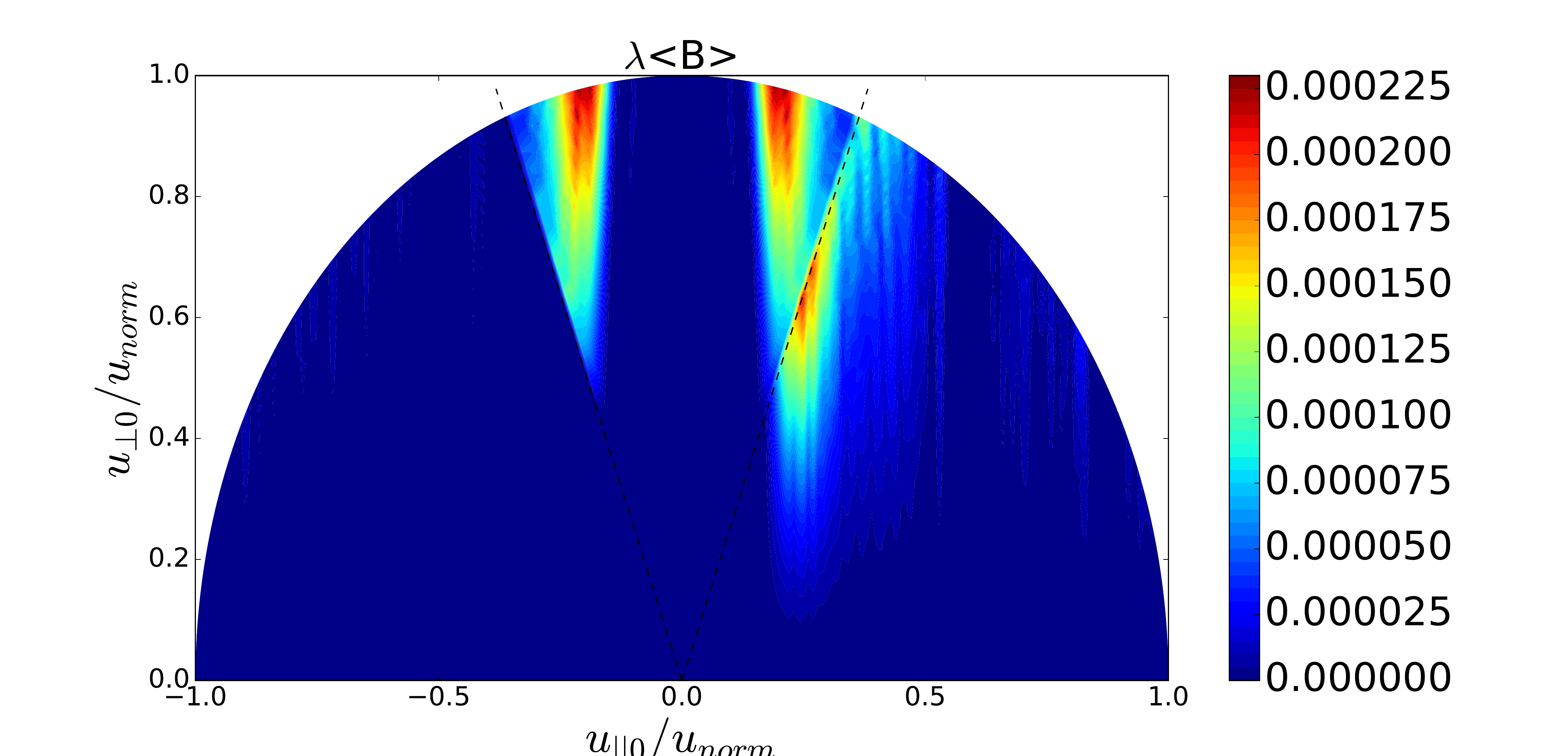}\\
(b)\\
\includegraphics[scale=0.2]{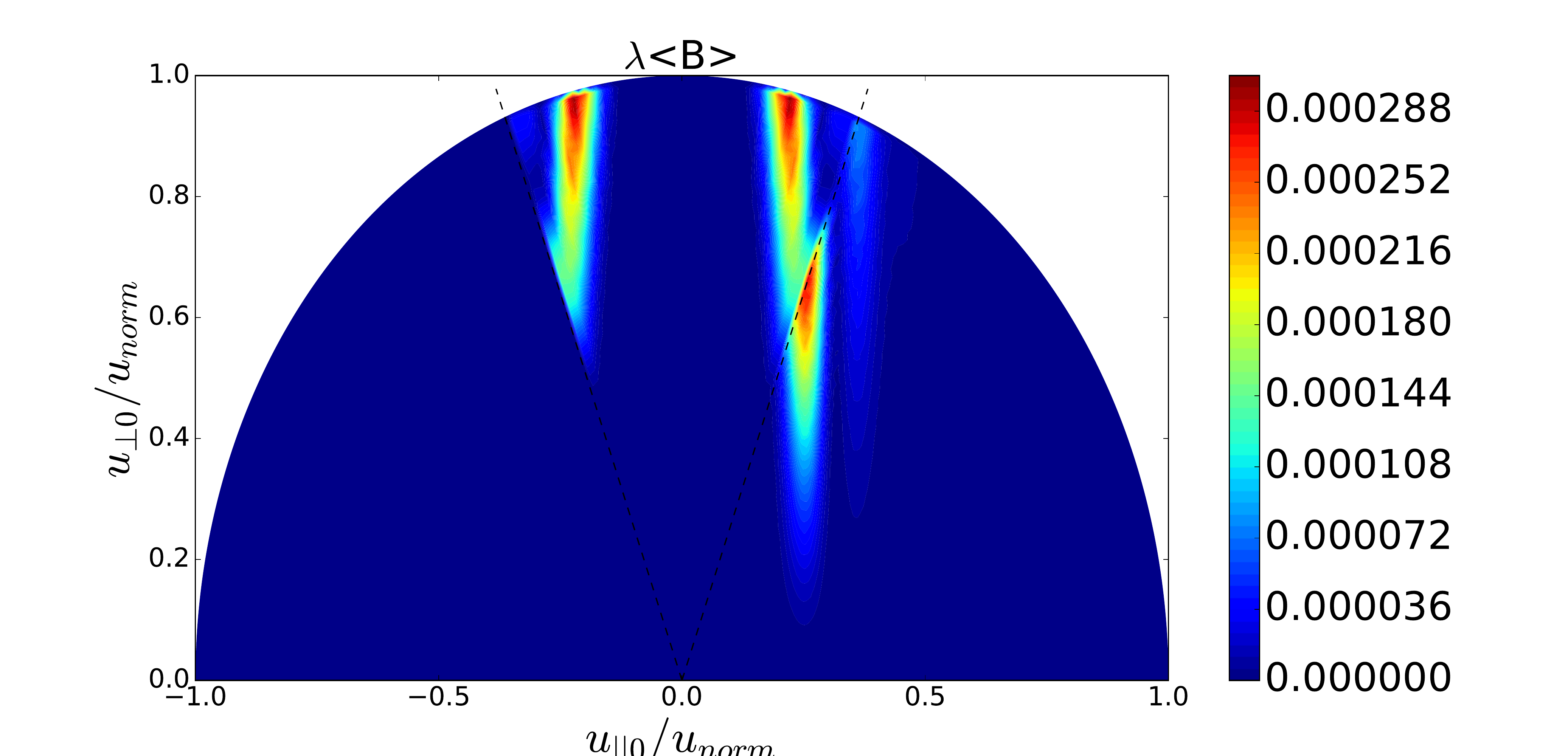}\\
(c)\\
\includegraphics[scale=0.2]{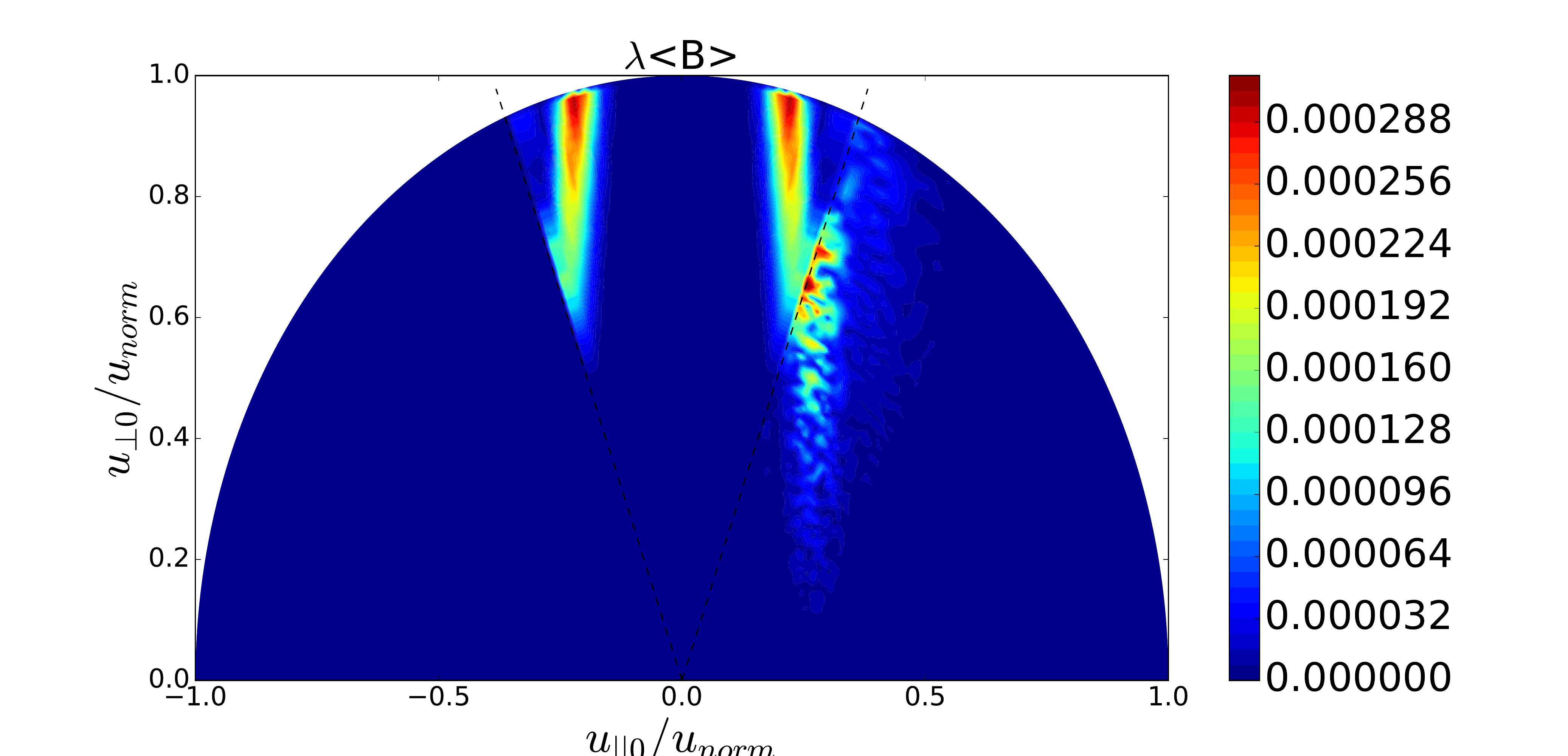}
\caption{ 2-D contour plots of the bounce-averaged quasilinear diffusion coefficient in velocity space $(u_{\|0}, u_{\perp0})$ at $r/a=0.25$ using (a) K-E form, (b) Positive-definite form with $\theta_{ref}=-\pi$, and (c) Positive-definite form with $\theta_{ref}=0$. The contours in (a), (b), and (c) correspond to the power absorption in green, red, and cyan curves of Figure 5, respectively. Here $u_{norm}$ is the momentum corresponding to the energy of He3 500KeV. The dashed lines are the trapped-passing boundaries, and the unit of the  $\lambda \langle B\rangle$ is $v_{norm}^{4}$ where $v_{norm}$ is the speed corresponding to $u_{norm}$ \cite{Harvey:IAEA1992}. The simulation domain in this plot is inside the circle of the radius=1.0.} 
\end{figure*}

 \begin{figure*}
(a)\\
\includegraphics[scale=0.2]{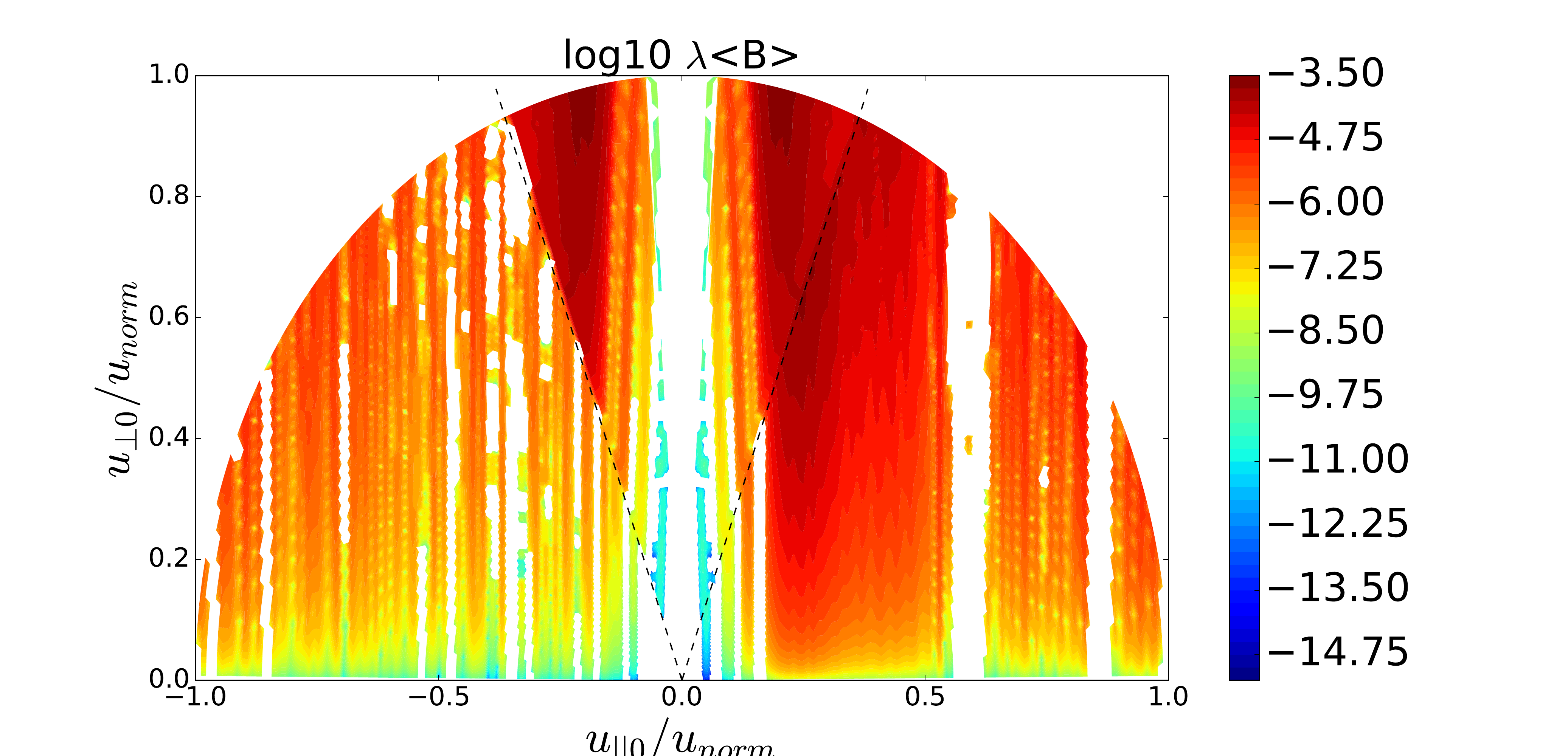}\\
(b)\\
\includegraphics[scale=0.2]{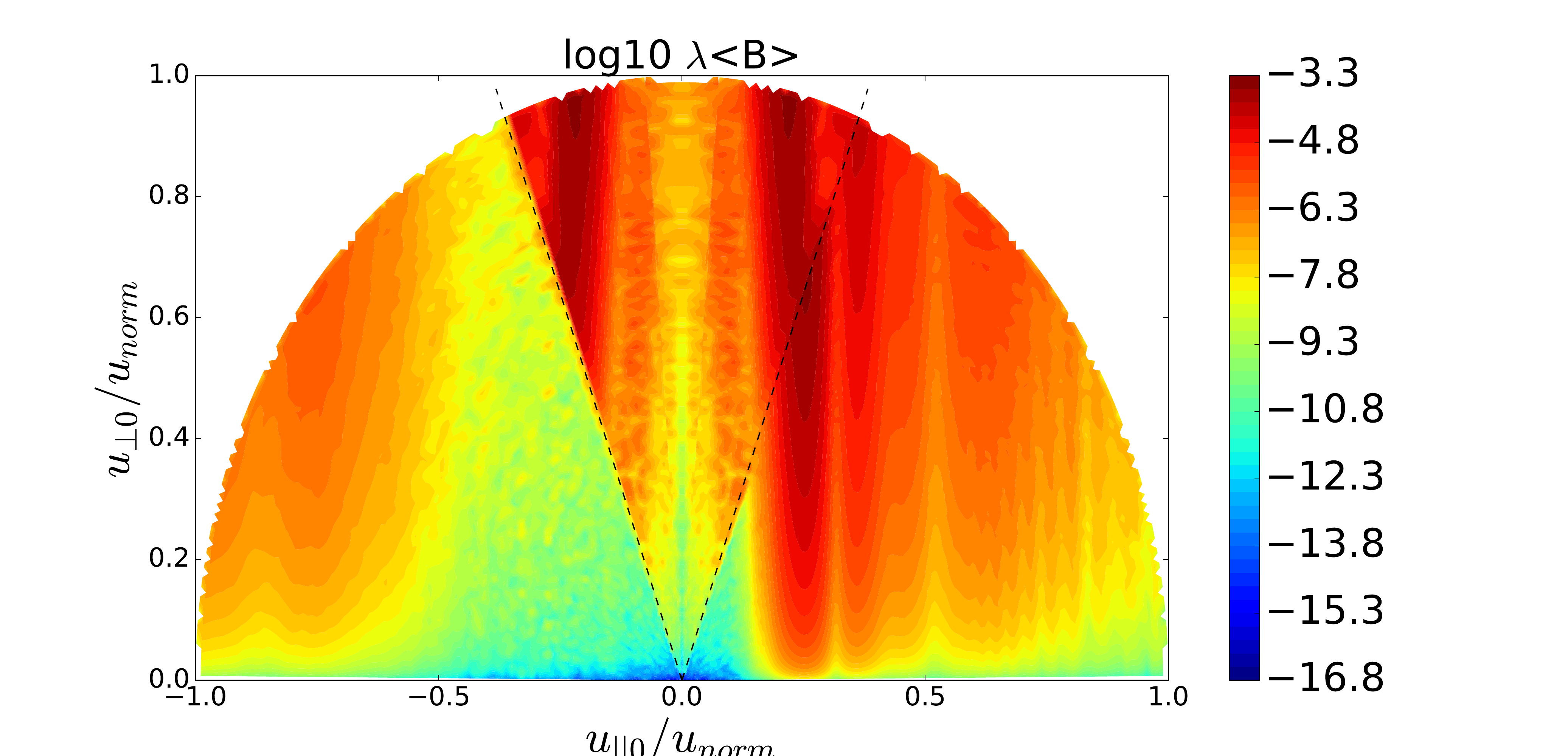}\\
(c)\\
\includegraphics[scale=0.2]{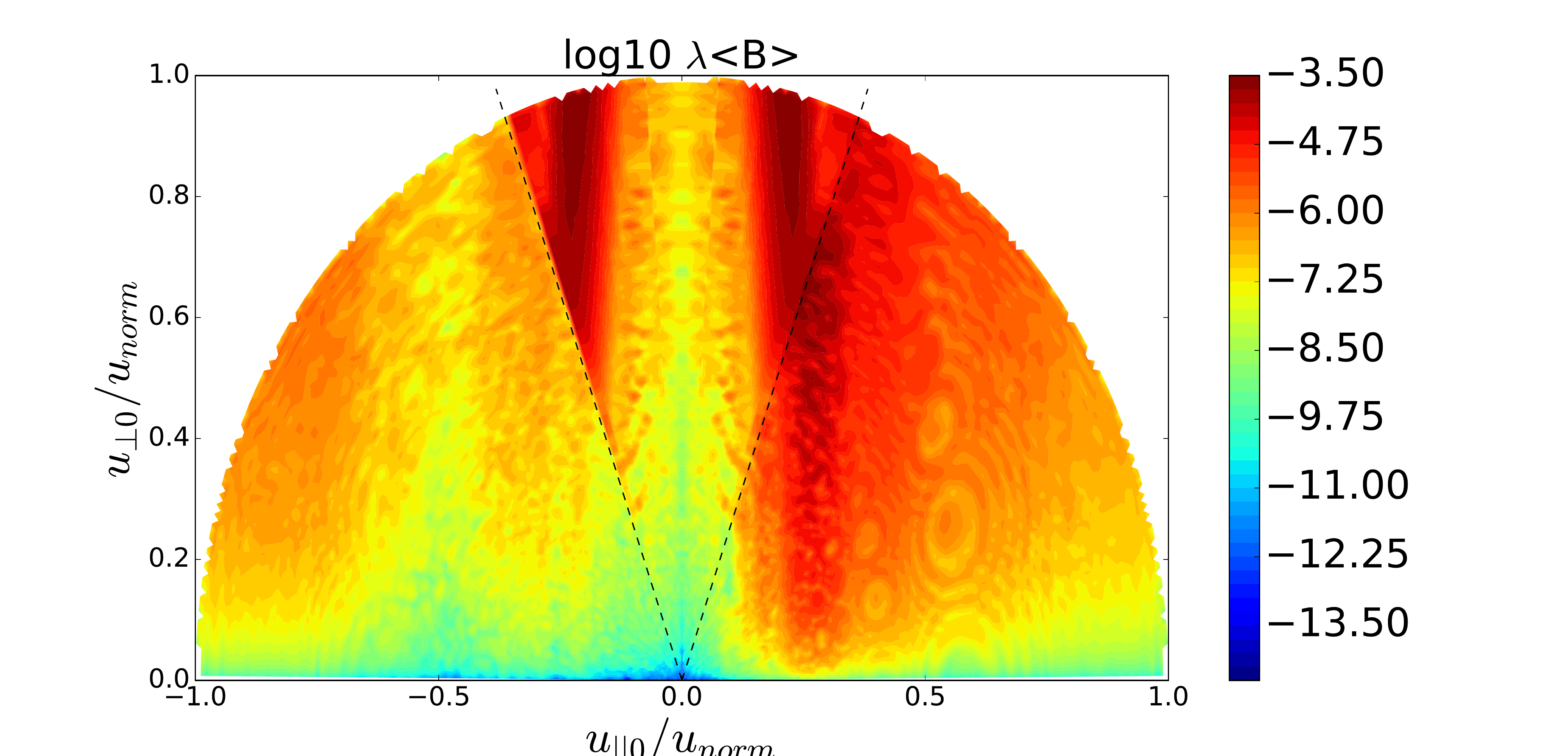}
\caption{The same 2-D contour plots as Figure 6 in log scale.} 
\end{figure*}

\subsection{Non-resonant interactions}

The additional contribution by the non-resonant interaction from $N(t_{nr})$ in Eq. (\ref{Nt}) results in the continuous evaluation of the integrand of Eq. (\ref{Dql5}) in both $\mathbf{k}$ and $\mathbf{v}$ space. Because the resonance condition depends on both $k_\|$ and $v_\|$, including only the resonance interaction results in the discontinuity depending whether there exists a resonance or not. In the K-E diffusion form of Eq. (\ref{Dql2}), the resonance condition depends on only $k_{\|1}$ and the integral in $\mathbf{k_2}$ space does not depend on the resonance condition, so the discontinuity in $\mathbf{k}$ space by the resonance condition is not problematic in the K-E evaluation. On the other hand, the continuous integrand of Eq. (\ref{Dql5}) in $\mathbf{k}$ space is necessary in the positive-definite form to include the interference between the different spectra. 

In Figure 7-(a), the discontinuity in $\mathbf{v}$ space by ignoring the non-resonant interaction in the K-E form is shown in the large white blocks (e.g. inside $-0.1<u_{\|0}/u_{norm}<0.1$), while they are filled by the continuous values in the positive-definite form in Figure 7-(b) and (c) as well as the measured diffusion in Figure 7-(d). If we ignore $N(t_{nr})$ in the positive-definite form of Eq. (\ref{Dql5}) when there is no resonance for all $\mathbf{k}$ spectra at a certain velocity space grid, it results in the contours of Figure 8 for the positive-definite form, which also shows the similar white regions around $-0.1<u_{\|0}/u_{norm}<0.1$ as Figure 7-(a). Even in this case of Figure 8, at each velocity coordinate, $N(t_{nr})$ is included if there is at least one resonance in a $\mathbf{k}$ spectrum, because the continuous integrand of Eq. (\ref{Dql5}) in $\mathbf{k}$ space is necessary as explained in the previous paragraph. The change in the total power absorption by excluding $N(t_{nr})$ is small in this example: only $1\%$ change of total power absorption by $N(t_{nr})$ in Figure 7-(b) compared to the power absorption in Figure 8.


 \begin{figure*}
\includegraphics[scale=0.2]{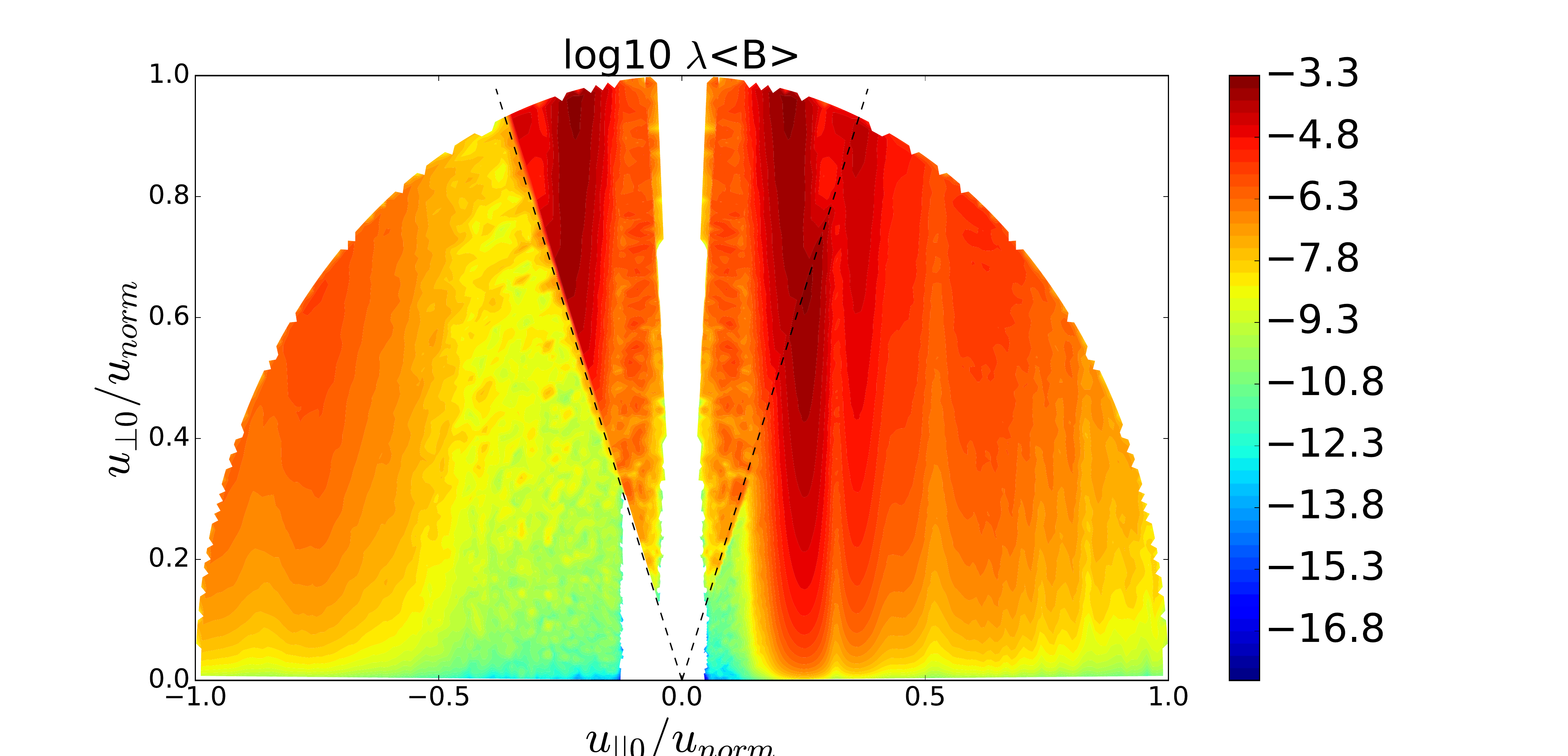}

\caption{The same 2-D contour plots as Figure 7-(b) for the positive-definite form with $\theta_{ref}=-\pi$ but ignoring non-resonant interaction $N(t_{nr})$ when there is no resonance in all $\mathbf{k}$ spectra} 
\end{figure*}

  \subsection{Evaluation periods}\label{sec4.3}
 Figure 9 shows the power profiles for the same case as Figure 5 by $\langle \mathbf{E}\cdot \mathbf{J} \rangle_w$ and $\dot{W}$ using the average of the bounce-averaged quasilinear diffusion coefficient in many poloidal periods. Note again that $\langle \mathbf{E}\cdot \mathbf{J} \rangle_w$ is evaluated by the plasma dispersion function in the homogeneous limit, so the power profiles by $\dot{W}$ with the positive definite form for the toroidal geometry do not necessarily match with $\langle \mathbf{E}\cdot \mathbf{J} \rangle_w$. As explained in Sec. \ref{sec3.2}, the non-periodic $\Phi_n$ results in different results depending on the number of periods. Figure 9-(a) and (b) show the changes for $\theta_{ref}=-\pi$ and $\theta_{ref}=0$, respectively.  As expected in Sec. \ref{sec:4.1}, the difference depending on the initial value of the evaluation range (poloidal reference) decreases as the number of evaluation periods increases (compare the yellow curves and the violet curves in Fig. 8). Although the difference between $\theta_{ref}=-\pi$ and $\theta_{ref}=0$ converges to zero, Figure 9 shows that the changes in many periods does not converge by increasing number of evaluation periods (see the changes between the red and violet graphs). The unexpected additional diffusion in many periods (especially at $r/a > 0.4$) is due to the missing decorrelation in many periods for a specific condition, when $mod(\Delta \Phi_n, 2\pi)=0$. If a poloidal spectral mode  satisfies the condition $mod(\Delta \Phi_n, 2\pi)=0$ for a velocity space grid at a flux surface, all other poloidal spectral modes satisfy the condition because $\Delta \Phi_n (m+1)=\Delta \Phi_n (m)+2\pi$. This unphysical diffusion can be reduced by adding toroidal mode summation or any decorrelation mechanism in the evaluation.  By adding the realistic decorrelation to the simulations, we may recover the perfect decorrelation assumption used when deriving Eq. (8) for the positive-definite form in Section 3.1. 
 
The possible decorrelation mechanisms in many poloidal periods are the collisions, the radial drift, and the perturbed orbit by the RF waves. For this example of ITER with $T_e=T_i=25 KeV$, $n_e=10^{20} /m^3$, $q=1.5$, and $R=6 m$, the collisional time by $\tau_i\simeq{1}/{\nu_{ie}}=3.44\times 10^5 \sqrt{(m_i/m_e)} T_e^{3/2}/(n_e \log\Lambda)\simeq 10^{-2}$ [sec] is much longer than the time for the one poloidal orbit $L_\|/v_{ti}\simeq 2\pi q R/v_{ti}\simeq 10^{-5}$ [sec]. However, the collisional decorrelation could be effective for ion cyclotron damping.  The previous studies \cite{Becoulet:PFB1991,Kasilov:NF1990} showed that the collisional decorrelation can make a significant impact on the cyclotron resonance in the inhomogeneous magnetic fields by including the collisional diffusion in the phase.  Additionally, a small change of the velocity due to the RF waves after each period can decorrelate it. The implementation of the decorrelation mechanism in the diffusion form will be investigated in the future.
 
 \begin{figure*}
(a) \;\;\;\;\;\;\;\;\;\;\;\;\;\;\;\;\;\;\;\;\;\;\;\;\;\;\;\;\;\;\;\;\;\;\;\;\;\;\;\;\;\;\;\;\;\;\;\;\;\;\;\;\;\;(b) \;\;\;\;\;\;\;\;\;\;\;\;\;\;\;\;\;\;\;\;\;\;\;\;\;\;\;\;\;\;\;\;\;\;\;\;\;\;\;\;\;\;\;\;\;\;\;\;\;\;\;\;\;\;\;\;\;\;\;\;\;\;\;\\
\includegraphics[scale=0.4]{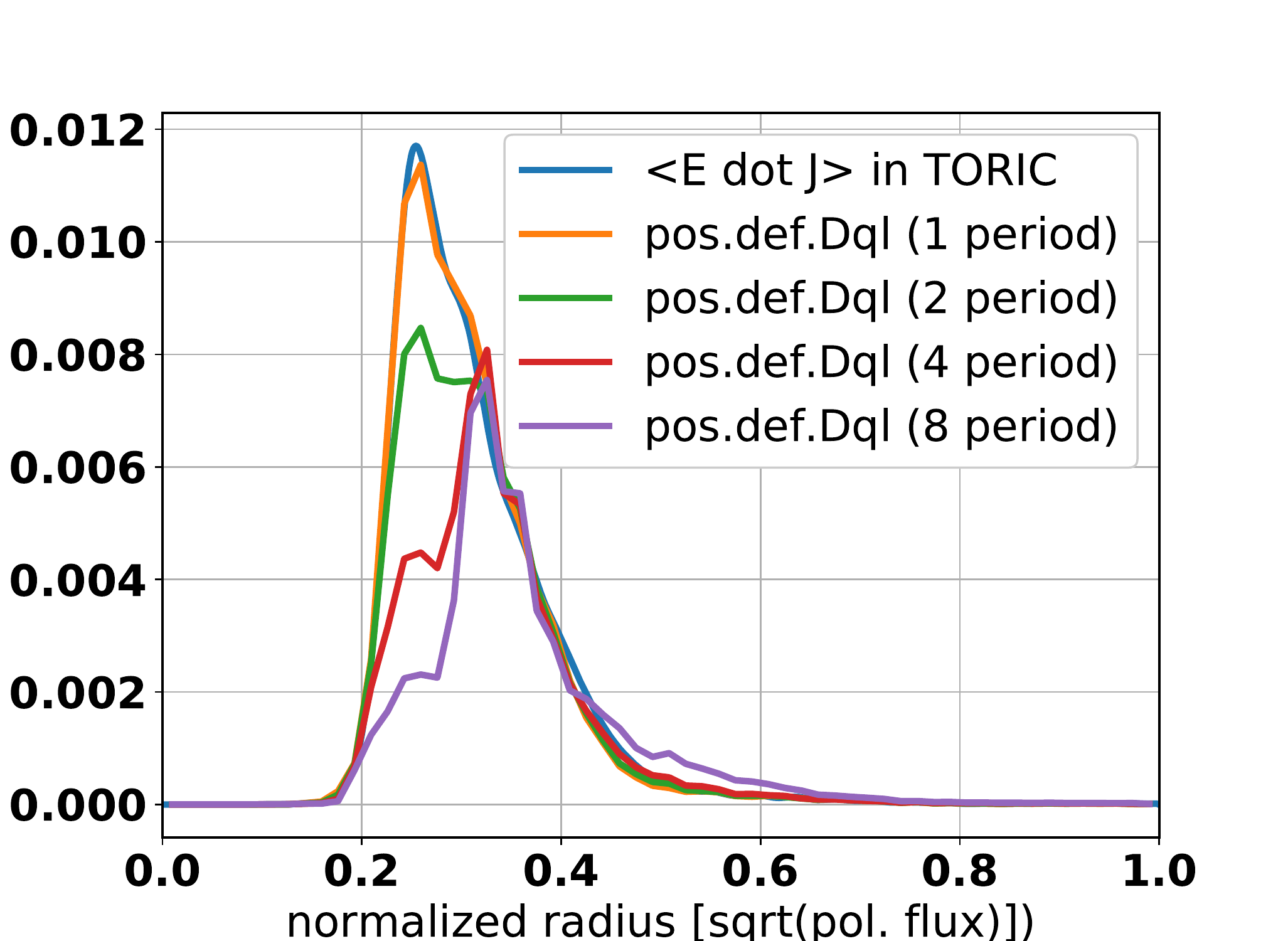}\includegraphics[scale=0.4]{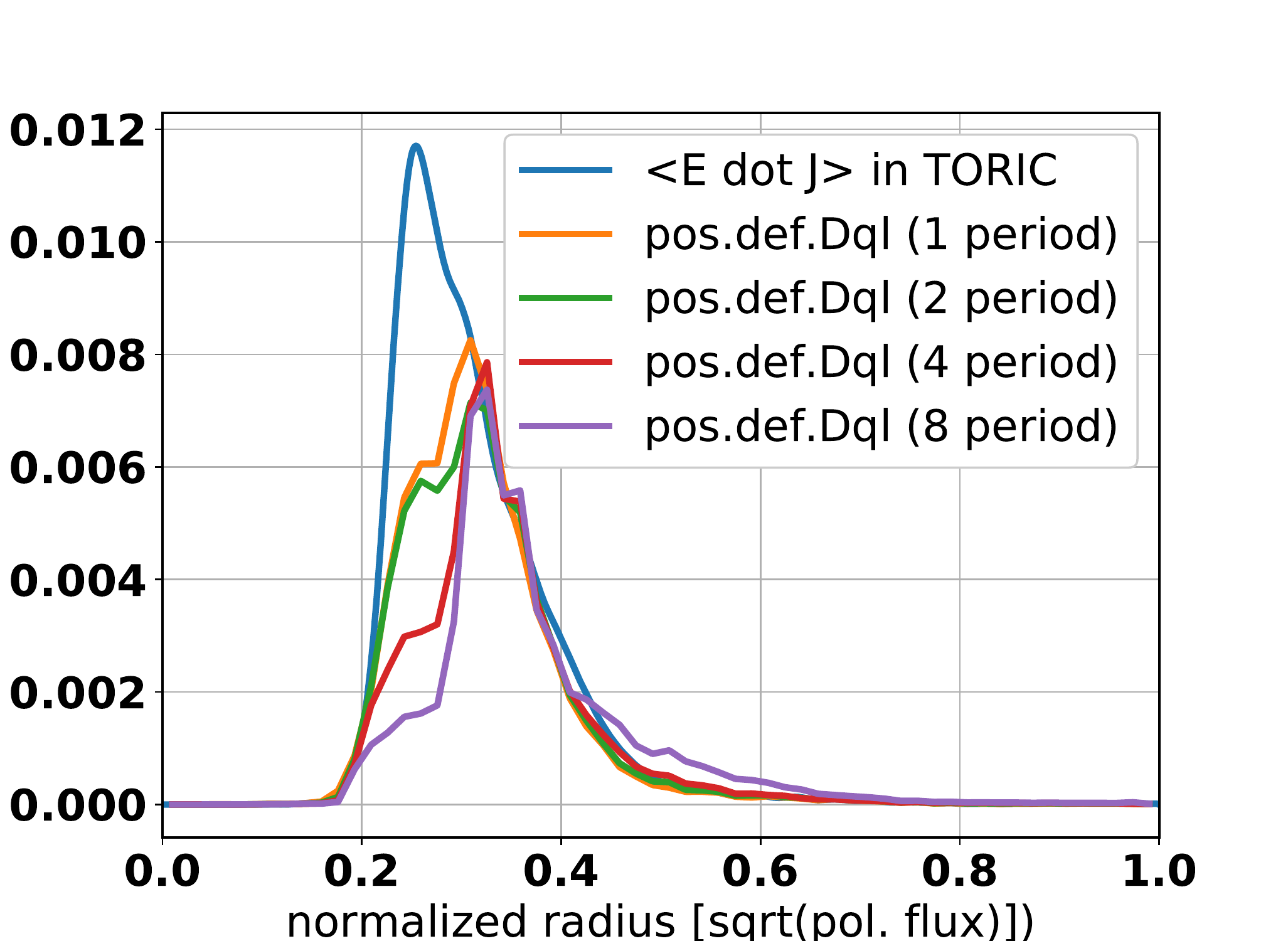}
\caption{Radial profiles of power absorption of the same case in Figure 5 with the evaluation in many poloidal periods for (a) $\theta_{ref}=-\pi$ and (b) $\theta_{ref}=0$} 
\end{figure*}

\section{Discussion}\label{sec:5}

In this paper, we derive the positive-definite form of bounce-averaged quasilinear diffusion coefficients in a toroidal geometry. Using the K-E diffusion coefficient cannot guarantee the positive definiteness of its bounce-average because of the broken symmetry between the trajectory integral and the bounce integral. By evaluating the trajectory integral using Airy functions in the toroidal geometry, we can obtain the positive-definite form in Eq. (\ref{Dql5}). As expected, using the positive-definite form reduces the numerical errors due to the negative diffusion in K-E form as shown in Figure 5. 

Furthermore, the positive-definite form can include other important toroidal effects that are ignored in the K-E form. The toroidal effects occur significantly around the inner-midplane, outer-midplane, and trapping tips for $d^2\Phi_n/dt^2\simeq 0$ when the wave-particle correlation length becomes so large that the parallel variation needs to be considered in the correlation length. Due to the long correlation length, the consecutive resonances in this region can be correlated with each other. Analytically, it can be shown in the cancellation between $\Phi_n$ and $\alpha(x_1t_++t_+^3/3)$ in $R(t_r)$ in Eq. (\ref{cancel1}) for the minimized phase mixing, which cannot be shown in the Dirac-delta function of the K-E form. Numerically, it can result in different results depending on the choice of the poloidal reference for the evaluation range. The evaluation range can be adjusted to provide the continuous evaluation to capture the correlations between resonances, as shown in Figure 5. The non-resonant contribution in the region of the long correlation length can be significant if $|d\Phi_n/dt|$ is sufficiently small, and the additional contribution makes the continuous diffusion in both $\mathbf{k}$ and $\mathbf{v}$ space. 

In spite of the good capability to capture some toroidal effects in the diffusion, the positive-definite form of this paper still cannot include all toroidal effects. First, in our form, the parallel variation is considered by the expansion around the resonance or non-resonant locations up to third order. If the variation in the correlation length cannot be described by the third order expansion, the form derived here fails to include the effect. For example, in a small radius flux surface, the non-resonant contribution from the inner-midplane and the outer-midplane can be mixed, while in our form they are independent of each other and some of their contributions may be double-counted.  On the other hand, some previous studies \cite{Becoulet:PFB1991, Lamalle:PPCF1997, Eester:PPCF2005} found the over-accentuation of the tangent resonance as the result of the trajectory integral and they suggested some methods to address the problems.  As the rapid decorrelation assumption ensuring a positive definite kick clashes with the phase stationarity close to the turning points, B{\'e}coulet suggested to simply omit the oscillatory part of the Airy function hereby ensuring a smooth connection between usual and higher order stationary phase points \cite{Becoulet:PFB1991}. Similarly, and based on a detailed assessment of Kaufman's ideas, Lamalle suggested to symmetrize the dielectric response by replacing the $k_\|$ in the Fried-Conte function by its average for the poloidal modes of the electric field and RF current density  \cite{Lamalle:PPCF1997}. For our treatment, our primary concern is likewise the symmetry of the result. 

Additionally, as explained in the introduction, the magnetic and $\nabla B$ drifts giving the finite orbit width can have a significant impact on the diffusion of the energetic ions. In this paper, the drifts are just ignored in the phase and the trajectory integrals for simplicity.  Because of the complexity in the diffusion in the mixed real and velocity phase space when the canonical toroidal angular momentum $p_\varphi$ is considered \cite{Kaufman:PF1972,Eriksson:PoP1994,Eriksson:PoP2005}, the numerical evaluation of the drift effects will be complicated.   Using the Monte-Carlo operator for the diffusion in $p_\varphi$, some particle codes have shown the radial transport \cite{Johnson:NF2006,Jucker:CPC2011,Hellsten:2004}, and the current and ion toroidal rotation drive \cite{Hellsten:1995,Murakami:2016} induced by RF waves. These topics will be the subject of future work to extend the analytical form of this paper to be used in the continuum Fokker-Planck code.  

Another problem of using the positive-definite form for the self-consistent solutions between Maxwell's equations and the Fokker-Planck equation is the lack of consistency with the plasma dispersion function used for the dielectric tensor, as explained in Section 3.3. Because the plasma dispersion function has the same assumption as that used for the K-E diffusion coefficients for the uniform phase along the parallel direction, it results in the mismatch of the assumption with the form for the parallel inhomogeneity. When the difference between the $ \langle \mathbf{E}\cdot \mathbf{J} \rangle_w$ by the plasma dispersion function and $\dot{W}$ of the positive definite form is small like Figure 5, the mismatch may have the negligible impact on the numerical convergence of the iteration between Maxwell's equation solver and Fokker-Planck equation solver. Otherwise, we may need to investigate the equivalent dielectric tensor to the positive-definite form using the trajectory integrals.  Because the Airy function in Eq. (\ref{Rt}) for the trajectory integral is a complex number (not pure real or imaginary) unlike the Dirac-delta function, both Hermitian part and anti-Hermitian part of the dielectric tensor need to be corrected according to the trajectory integral. 

The particle codes to measure the diffusion numerically \cite{Johnson:NF2006,Harvey:RF2001,Jucker:CPC2011,Hellsten:2004} can include the perturbed orbit effects as well as the toroidal effects. They can show more realistic diffusion for certain particles in a phase velocity (e.g. particle of the stagnation point about $v_\|\simeq0$), which cannot be captured easily in the analytical form as this paper. Nevertheless, the analytical form is also useful to unveil the hidden structures of the numerical results as given in the cancellation of Eq. (\ref{cancel1}). Furthermore, the comparisons between the results from analytical form and the particle code will be beneficial to increase the reliability of the theory.

 \newpage

\appendix
 \section{Relation between the conductivity tensor and the quasilinear diffusion tensor}
   In the Maxwell's equation,
\begin{eqnarray}
\nabla \times  {\nabla} \times \mathbf{E}+ \frac{\omega^2}{c^2}\mathbf{E}= -i \omega \mu_0 ( \mathbf{J}_p+\mathbf{J}_{ant})\ \label{Eq1},  
\end{eqnarray}
where $ \mathbf{J}_p$ is the plasma current and $ \mathbf{J}_{ant}$ is the external current at the antenna. The conductivity kernel tensor $\bar{\sigma}$ is defined by
        \begin{eqnarray}
 \mathbf{J}_p (\mathbf{r}) =\int d\mathbf{k}   \bar{\sigma} (\mathbf{r},\mathbf{k}) \cdot  \mathbf{E} ( \mathbf{k})e^{i \mathbf{k}\cdot\mathbf{r}} ,
  \end{eqnarray}

 For the kinetic description of the wave, the Maxwell's equation is coupled with the Fokker-Planck equation for a species $s$,
  \begin{eqnarray}
\frac{\partial f_s}{\partial t} +(\mathbf{v_D}+v_\| \mathbf{b}) \cdot \nabla f_s  = C(f_s)+Q(f_s) \label{FP},
\end{eqnarray}
where  $\mathbf{v}_{D}$ is the drift velocity and $C(f_s)$ is the Fokker-Planck collision operator. The quasilinear velocity diffusion $Q(f_s)$ can be obtained by taking an average of the product of two oscillating quantities,
\begin{eqnarray}
Q(f_s)&=&-\frac{q}{m}\left \langle \nabla_v \cdot \left[\left({\mathbf{E}} +\frac{\mathbf{v} \times {\mathbf{B}} }{c}\right) \widetilde{f}_s \right] \right \rangle_w \label{Q0},
\end{eqnarray}
where  ${ \widetilde{f_s}}$ is the weakly perturbed distribution function by RF waves, and $\langle ...\rangle_w$ is the average in a sufficiently long time and length scale compared to the wave oscillation. The conductivity tensor can be defined by the perturbed distribution using $\mathbf {J}=\sum_s \int d\mathbf{v} q_s \mathbf{v} \widetilde{f}_s$.

By taking bounce-average of the equation after ignoring the drift term, the Fokker-Planck equation can be
  \begin{eqnarray}
\frac{\partial f }{\partial t} +\langle Q(f) \rangle_b =  \langle C(f) \rangle_b\label{FP},
\end{eqnarray}
where $f$ can be defined by the invariant velocity variables $\mathbf{v_c}$ in the bounce time. 

   In many studies \cite{smithe1989local,Jaeger:NF1988,Brambilla:1988PPCF}, the relation between the conductivity tensor and the quasilinear diffusion has been investigated. The Poynting theorem (e.g. dot-product of Eq. (\ref{Eq1}) with the electric fields) show the relation,
   \begin{eqnarray}
 \langle \mathbf{J}\cdot \mathbf{E} \rangle_w =\dot{W} + \nabla \cdot \mathbf{T}  \label{rel1},
\end{eqnarray}
where $\mathbf{T}$ is the kinetic flux and $\dot{W}$ is the power absorption that can be defined by the quasilinear linear diffusion
   \begin{eqnarray}
   \dot{W} = \int d\mathbf{v} \frac{mv^2}{2} Q(f). \label{Wdot1}
   \end{eqnarray}
    
    The volume integral of Eq. (\ref{rel1}) shows the relation more clearly because the kinetic flux vanishes. The volume integral of the Joule heating term is
        \begin{eqnarray}
\fl  \frac{1}{2} Re\left[ \int d\mathbf{r} ( \mathbf{J}\cdot \mathbf{E}^* )\right] &=& \frac{1}{2} Re\left[ \int d\mathbf{r} \int d\mathbf{k}_1 \int d\mathbf{k}_2  ( \mathbf{E}^*(  \mathbf{k}_2)  \cdot  \bar{\sigma} (\mathbf{r},\mathbf{k}) \cdot  \mathbf{E} ( \mathbf{k}_1)) e^{i( ( \mathbf{k}_1- \mathbf{k}_2) \cdot \mathbf{r})}\right] \label{jdote1}\\
\fl  &=& \int d\mathbf{r} \int d\mathbf{v} \frac{mv^2}{2} Q(f). \label{Wdot2}
  \end{eqnarray}
  Because $ \bar{\sigma}$ has the velocity integral $\int d\mathbf{v}$, Eq. (\ref{jdote1}) has $\int d\mathbf{r} \int d\mathbf{v}$, which is an integral over cyclical motions and constants of motion of the phase space. The remaining integrands in Eq. (\ref{jdote1}) are related to the quasilinear diffusion in Eq. (\ref{Wdot2}).

\section{quasilinear diffusion coefficient in a homogenous magnetic field }\label{sec:A1}
In a uniform magnetic field with spatially uniform plasmas, the Kennel-Engelmann quasilinear diffusion operator  \cite{Kennel:POF1966} can be defined by
\begin{eqnarray}
Q(f)&=&-\frac{q}{m}\left \langle \nabla_v \cdot \left[\left(\mathbf{E} +\frac{\mathbf{v} \times \mathbf{B} }{c}\right) {f} \right] \right \rangle_w \label{Q1}\\
&\simeq&-\frac{q}{m}\nabla_v \cdot \left[ \sum_\mathbf{k} \left \{\stackrel{\leftrightarrow}{I} \left(1-\frac{\mathbf{k} \cdot \mathbf{v}}{\omega} \right) +\frac{\mathbf{k} \mathbf{v}}{\omega} \right\}\cdot\mathbf{E}_\mathbf{-k}   f_{\mathbf{k} } \right], 
 \label{Dql_0}
\end{eqnarray}
where $\stackrel{\leftrightarrow}{I}$ is the unit tensor. We have used the Fourier analyzed fluctuating electric field, $\mathbf{E}=\sum_\mathbf{k}  \mathbf{E}_\mathbf{k}  \exp(i\mathbf{k} \cdot \mathbf{r}-i \omega_\mathbf{k} t)$, the fluctuating magnetic field $\mathbf{B}=\sum_\mathbf{k} \mathbf{B}_\mathbf{k}  \exp(i\mathbf{k} \cdot \mathbf{r}-i \omega_\mathbf{k} t)$, and the fluctuating distribution function, $f=\sum_\mathbf{k} f_\mathbf{k}  \exp(i\mathbf{k} \cdot \mathbf{r}-i \omega_\mathbf{k} t)$. The functions $\mathbf{E}_\mathbf{k} \equiv   \mathbf{E}(\omega_{\mathbf{k} },\mathbf{k})$, $\mathbf{B}_\mathbf{k} \equiv   \mathbf{B}(\omega_{\mathbf{k} },\mathbf{k})$, and $f_\mathbf{k} \equiv  f(\omega_{\mathbf{k} },\mathbf{k})$ satisfy the relation $f_{-\mathbf{k} }\equiv f(\omega_{-\mathbf{k} },-\mathbf{k})=f^*(\omega_{\mathbf{k} },\mathbf{k})$ where $*$ denotes complex conjugate and $\omega_{\mathbf{k} }=-\omega^*_{-\mathbf{k} }$. Faraday's law has been used in going from (\ref{Q1}) to (\ref{Dql_0}) to write $\mathbf{B}_\mathbf{k} =(c/\omega) \mathbf{k} \times \mathbf{E}_\mathbf{k} $. 
The quasilinear operator can be written as 
 \begin{eqnarray}
Q(f)\equiv \frac{q}{m}\left[\frac{1}{v_{\perp} }\frac{\partial}{\partial v_{\perp}} \left(v_{\perp}\Gamma_\perp\right) +\frac{1}{v_{\perp} }\frac{\partial \Gamma_\phi}{\partial \phi} +\frac{\partial \Gamma_\|}{\partial v_{\|}}\right].
 \label{Dql_1}
\end{eqnarray}
The flux in the perpendicular direction is
 \begin{eqnarray}
\Gamma_\perp= -  \sum_\mathbf{k}\bigg \{E^{*}_{\mathbf{k} ,\perp}\bigg(1-\frac{k_{\|}v_{\|}}{\omega} \bigg) +E^{*}_{\mathbf{k} ,\|} \frac{k_{\perp}v_{\|}}{\omega} \cos{(\phi-\beta)} \bigg \}f_{\mathbf{k} }
 \label{Gamma_perp}.
\end{eqnarray}
Here, the velocity is defined as
$\mathbf{v}=v_{\perp} \cos{\phi}\;\mathbf{e_x}+ v_{\perp} \sin{\phi} \;\mathbf{e_y}+v_{\|} \mathbf{e_\|}$, where $\phi$ is the gyro phase angle, and $x$ and $y$ are the orthogonal coordinates in the perpendicular plane to the static magnetic field. The wavenumber vector is defined as $\mathbf{k}=k_{\perp} \cos{\beta}\;\mathbf{e_x}+ k_{\perp} \sin{\beta}\;\mathbf{e_y}+k_{\|}\mathbf{e_\|}$.
The flux in the gyro-phase direction is
 \begin{eqnarray}
\Gamma_\phi&=& -  \sum_\mathbf{k}\bigg\{E^{*}_{\mathbf{k} ,\phi} \bigg(1-\frac{k_{\perp}v_{\perp}}{\omega} \cos{(\phi-\beta)}-\frac{k_{\|}v_{\|}}{\omega} \bigg)\nonumber \\
&-& E^{*}_{\mathbf{k} ,\perp}\frac{k_{\perp}v_{\perp}}{\omega}  \sin{(\phi-\beta)} -E^{*}_{\mathbf{k} ,\|} \frac{k_{\perp}v_{\|}}{\omega} \sin{(\phi-\beta)} \bigg \}{f}_\mathbf{k} ,
 \label{Gamma_alpha}
\end{eqnarray}
and the flux in the parallel direction is
 \begin{eqnarray}
\Gamma_\|=  - \sum_\mathbf{k}\bigg\{E^{*}_{\mathbf{k} ,\|} \bigg(1-\frac{k_{\perp}v_{\perp}}{\omega} \cos{(\phi-\beta)} \bigg) + E^{*}_{\mathbf{k} ,\perp}\frac{k_{\|}v_{\perp}}{\omega}\bigg \}{f}_\mathbf{k} .
 \label{Gamma_par}
\end{eqnarray}
Here, the perturbed fluctuating distribution function consistent with a single mode wave is
 \begin{eqnarray}
\fl f_\mathbf{k}  &=& - \frac{q}{m}e^{-i \mathbf{k} \cdot \mathbf{r} + i \omega t} \int^t_{-\infty} d t^\prime e^{i \mathbf{k} \cdot \mathbf{r}^{\prime} -i \omega t^{\prime}} \mathbf{E_\mathbf{k} } \cdot \left[\stackrel{\leftrightarrow}{I} \left(1-\frac{\mathbf{v}^\prime \cdot \mathbf{k}}{\omega} \right) +\frac{\mathbf{v}^\prime \mathbf{k}}{\omega} \right] \cdot  \nabla_{v^{\prime}} f ,  \label{f_k1}
\end{eqnarray}
where $(t^{\prime},\mathbf{r}^{\prime},\mathbf{v}^\prime)$ is a point of phase space along the zero-order particle trajectory. The trajectory end point corresponds to $(t,\mathbf{r},\mathbf{v})$. The background distribution, $f=f(t,\mathbf{r},v_{\perp}, v_{\|})$, is gyro-phase independent because of the fast gyro-motion. As a result,
 \begin{eqnarray}	
f_\mathbf{k}  &=&-\frac{q}{m} \int_0^{\infty} d\tau \exp({i\alpha})  \bigg\{ \cos{(\eta+\Omega \tau)}(( E_{\mathbf{k} ,+}+E_{\mathbf{k} ,-})U-E_{\mathbf{k},\|}V)\nonumber\\
&&-i\sin{(\eta+\Omega \tau)}( E_{\mathbf{k} ,+}-E_{\mathbf{k} ,-})U +E_{\mathbf{k} ,\|} \frac{\partial f}{\partial v_{\|}}\bigg \}. \label{f_k2}
\end{eqnarray}
Here, $\tau=t-t^{\prime}$, and $\alpha=(\omega-k_{\|}v_{\|})\tau -\lambda(  \sin{(\eta+\Omega \tau)}- \sin{(\eta)}) $, where $\lambda= {k_{\perp}v_{\perp}}/{\Omega}$ and $\eta=\phi-\beta$. Also, $U={\partial f}/{\partial v_{\perp}}+ ({k_{\|}}/{\omega})\left( v_{\perp}{\partial f}/{\partial v_{\|}}-v_{\|}{\partial f}/{\partial v_{\perp}}\right)$, and $V= ({k_{\perp}}/{\omega})\left( v_{\perp}{\partial f}/{\partial v_{\|}}-v_{\|}{\partial f}/{\partial v_{\perp}}\right)$. We follow Stix' notation \cite{Stix:AIP1992}.

For the energy transfer, the contribution of the flux in the gyro-phase direction vanishes due to the integral over $\phi$. Using the Bessel function expansion for the sinusoid phase, 
 \begin{eqnarray}
e^{i\lambda \sin {\eta}}&=&\sum_n e^{in\eta} J_n(\lambda),\nonumber\\
\sin{\eta} e^{i\lambda \sin {\eta}}&=&-\sum_n i e^{in\eta} J_n^{\prime}(\lambda),\nonumber\\
\cos{\eta} e^{i\lambda \sin {\eta}}&=&\sum_n \frac{n}{\lambda} e^{in\eta} J_n(\lambda),\label{Bess00}
\end{eqnarray}
the gyro-averaged quasilinear diffusion  \cite{Kennel:POF1966,Stix:AIP1992} is
\begin{eqnarray}
Q(f)&=&\frac{{\pi} q^2 }{m^2}    \sum_n G \bigg (v^2_{\perp} \delta (\omega-k_{\|}v_{\|}-n\Omega) |\chi_{\mathbf{k},n}|^2 G(f) \bigg ) \label{P_abs1}\label{P_abs2}
\end{eqnarray} 
where $\chi_{\mathbf{k} ,n}= E_{\mathbf{k} ,+} J_{n-1}/\sqrt{2} +E_{\mathbf{k} ,-} J_{n+1}/\sqrt{2}+({v_{\|}}/v_{\perp})E_{\mathbf{k} ,\|} J_{n}$ is the effective electric field, and the operator $G$ is 
\begin{eqnarray}
G(f)=\left(1-\frac{k_{\|}v_{\|}}{\omega}\right)\frac{1}{v_{\perp}}\frac{\partial f}{\partial v_{\perp}}+\frac{k_{\|}v_{\perp}}{\omega}\frac{1}{v_{\perp}}\frac{\partial f}{\partial v_{\|}} \label{KE_diff}
\end{eqnarray} 

The quasilinear diffusion coefficient can be defined by
\begin{eqnarray}
Q(f)&\equiv& \nabla_\mathbf{v} \cdot D_{ql} \cdot \nabla_\mathbf{v} f \label{Qapp},
\end{eqnarray}
where the coefficient tensor is given by
\begin{eqnarray}
 \fl  D_{ql} =  \frac{{\pi} q^2 }{m^2}   \sum_n   \left( (\mathbf{P_n}(\mathbf{k})\cdot\mathbf{E(\mathbf{k})})\mathbf{G}(\mathbf{k})\right)^* \delta(\omega-k_\|v_\|-n\Omega) \left( (\mathbf{P_n}(\mathbf{k}) \cdot\mathbf{E(\mathbf{k})})\mathbf{G}(\mathbf{k})\right).  \label{Dql1app}
 \end{eqnarray}
 The polarization vector $\mathbf{P_n}$ and the diffusion direction vector $\mathbf{G}$ are determined by the effective potential $\chi_{\mathbf{k} ,n}$ and the operator $G$,
   \begin{eqnarray}
  \mathbf{P_n} \cdot \mathbf{E}&=&\chi_{\mathbf{k} ,n} , \label{KE_P1}\\
  \mathbf{G}&=&\left(1-\frac{k_{\|}v_{\|}}{\omega}\right)\hat{\mathbf{v}}_\perp+\frac{k_{\|} v_{\perp}}{\omega}\hat{\mathbf{v}}_\|,\label{KE_G1}
   \end{eqnarray}
   
   The Dirac-delta function is obtained by the trajectory integral for
   \begin{eqnarray}
 \fl  \int_{-\infty}^t dt^{\prime} A(k)   e^{i\int_{t\prime}^t dt^{\prime\prime}(\omega-n\Omega-k_{\|}v_\|)}&=&   -\int_{-\infty}^t dt^{\prime} \frac{A(k)}{i(\omega-n\Omega-k_{\|}v_\|)}  \label{delta1}\\
  &\rightarrow& \pi A(k) \delta(\omega-k_\|v_\|-n\Omega) \label{delta2}
     \end{eqnarray}
\clearpage
\newpage

\section{Constants of Motion}

We declare that there exist three (momentum) constants of motion, $\mathbf{v_c} (\mathbf{r},\mathbf{v})$, and their Hamiltonian conjugate coordinates, $\mathbf{x_c} (\mathbf{r},\mathbf{v})$. The divergence theorem on the Jacobian will then hold,
     \begin{eqnarray}
\nabla_\mathbf{v_c} \cdot  \frac{\partial \mathbf{v_c}}{\partial \mathbf{v}} +\nabla_\mathbf{x_c} \cdot \frac{\partial \mathbf{x_c}}{\partial \mathbf{v}}=0,
   \end{eqnarray}
and so for the outer $\nabla_\mathbf{v_c}$ derivative (in the quasilinear equation), the chain rule, together with the divergence theorem gives
     \begin{eqnarray}
\nabla_\mathbf{v} \cdot \mathbf{\Gamma}&=&\frac{\partial \mathbf{v_c}}{\partial \mathbf{v}}^T: \nabla_\mathbf{v_c} \Gamma   +\frac{\partial \mathbf{x_c}}{\partial \mathbf{v}}^T: \nabla_\mathbf{x_c}\Gamma \nonumber \\
&=&\frac{\partial \mathbf{v_c}}{\partial \mathbf{v}}^T: \nabla_\mathbf{v_c} \Gamma   +\frac{\partial \mathbf{x_c}}{\partial \mathbf{v}}^T: \nabla_\mathbf{x_c} +\left(\nabla_\mathbf{v_c} \cdot  \frac{\partial \mathbf{v_c}}{\partial \mathbf{v}} +\nabla_\mathbf{x_c} \cdot \frac{\partial \mathbf{x_c}}{\partial \mathbf{v}}\right) \cdot \Gamma  \nonumber\\
&=&\nabla_\mathbf{v_c} \cdot \left(\frac{\partial \mathbf{v_c}}{\partial \mathbf{v}}^T \cdot \Gamma\right)   + \nabla_\mathbf{x_c} \cdot \left( \frac{\partial \mathbf{x_c}}{\partial \mathbf{v}}^T \cdot \Gamma\right) \label{AppGam},
   \end{eqnarray}
   where $\Gamma=-(q/m)(\mathbf{E}(\mathbf{r})+ \mathbf{v}\times\mathbf{B}(\mathbf{r}))^*f_1(\mathbf{r},\mathbf{v})$. In practice, this is only ever done to within a guiding center Hamiltonian, and so one typically has $\rho/R$ error terms in the divergence relation. 
   
   A favorite set of constants-of-motion for tokamaks is $\mathbf{v_c}=\{E,\mu,P_\varphi\}$, e.g., energy, magnetic moment, and canonical angular momentum, with $\mathbf{x_c}=\{t,\phi,\varphi_c\}$ as the conjugate coordinates.  It is also common in software programming to use $\mathbf{v_c}=\{v_{\perp0}, v_{\|0}, B/B_0\}$, e.g., perpendicular and parallel velocity and flux surface at the outboard midplane as constants-of-motion, with somewhat less obvious conjugate coordinates in that case.
We also declare that $f_0$ is a function of the constants of motion only, $\nabla_\mathbf{x_c} f_0=0$, so that for the inner $\nabla_\mathbf{v}$ derivative (in the perturbed distribution equation) we have
     \begin{eqnarray}
\nabla_\mathbf{v} f_0= \frac{\partial \mathbf{v_c}}{\partial \mathbf{v}}^T \cdot  \nabla_\mathbf{v_c} f_0,
   \end{eqnarray}
   and 
        \begin{eqnarray}
\fl f_1(\mathbf{r},\mathbf{v})= -\frac{q}{m}\int_{t_\infty}^{t} dt^{\prime} \left(\mathbf{E}(\mathbf{r_0}(t^{\prime}))+ \mathbf{v_0}(t^{\prime})\times\mathbf{B}(\mathbf{r_0}(t^{\prime})\right) \cdot \frac{\partial \mathbf{v_c}}{\partial \mathbf{v}}(\mathbf{r_0}(t^{\prime}),\mathbf{v_0}(t^{\prime}))\cdot \nabla_\mathbf{v_c} f_0(\mathbf{v_c}) \label{Appf1},
   \end{eqnarray}
   where the unperturbed orbits are $\mathbf{r_0}(t^\prime)$ and $\mathbf{v_0}(t^\prime)$, satisfying $\mathbf{r_0}(t)=\mathbf{r}$, $\mathbf{v_0}(t)=\mathbf{v}$, $\mathbf{r_\infty}=\mathbf{r_0}(t_\infty)$ and $\mathbf{v_\infty}=\mathbf{v_0}(t_\infty)$.
Note that since $\nabla_\mathbf{v_c} f_0(\mathbf{v_c})$ is constant along the trajectory, it can sit outside the $t^\prime$ integral, giving us the inner derivative of the diffusion equation.

 \section*{Acknowledgments}
 We would like to thank Dr. Peter Catto, Dr. Robert Harvey, and Dr. Yuri Petrov for their useful advice on this study. This work was supported by US DoE Contract No. DE-FC02-01ER54648 under a Scientific Discovery through Advanced Computing Initiative. This research used resources of MIT and NERSC by the US DoE Contract No.DE-AC02-05CH11231\\
 
\pagebreak

  \newpage

\begin{thebibliography}{100}
\bibitem{Stix:AIP1992}
  T. H. Stix, 1992 {\em Waves in Plasmas}
(AIP Press) ISBN 0-88318-859-7

\bibitem{Cary:PRL1990}
J. R. Cary and D. F. Escande and A. D. Verga 1990 {\em Physical Review Letters} {\bf 65} 3132

\bibitem{laval1999controversies}
 G. Laval and D. Pesme 1999 {\em Plasma physics and controlled fusion} {\bf 41} A239

\bibitem{Lee:POP2011}
J. P. Lee and P. T. Bonoli and J. C. Wright 2011 {\em Physics of Plasmas} {\bf 18} 012503

\bibitem{Kennel:POF1966}
C. F. Kennel and F. Engelmann 1966  {\em Physics of Fluids} {\bf 9} 2377--2388

\bibitem{Brambilla:PPCF1999}
M. Brambilla 1999 {\em Plasma Physics and Controlled Fusion } {\bf 41}  1
\bibitem{Jaeger:PoP2011}
E. F. Jaeger, L. A. Berry, E. DAzevedo, D. B. Batchelor, and M. D. Carter 2001 {\em Physics of Plasmas } {\bf 8} 1573



\bibitem{Brambilla:PLA1994}
M. Brambilla 1994 {\em Physics Letters A} {\bf 188}  376

\bibitem{smithe:PRL1988}
D. Smithe, P. Colestock, T. Kammash, and R. Kashuba 1988 {\em Physics Review Letters} {\bf 1988} 801 


\bibitem{Berry:PoP2016}
L. A. Berry, E. F. Jaeger, C. K. Phillips, C. H. Lau, N. Bertelli, and D. L. Green  2016 {\em Physics of Plasmas } {\bf 23} 102504

\bibitem{Bernstein:PF1981}
Bernstein I.B. and Baxter D.C. 1981 {\em Phys. Fluids } {\bf 24} 108-26 

\bibitem{Kerbel:PF1985}
 Kerbel G.D. and McCoy M.G. 1985 {\em Phys. Fluids } {\bf 28} 3629-53 
 
\bibitem{Becoulet:PFB1991} 
B{\'e}coulet A., Gambier D.J. and Samain A. 1991 {\em Phys. Fluids B } {\bf 3} 137-50

\bibitem{Catto:PF1992}
Catto P.J. and Myra J.R. 1992 {\em  Phys. Fluids B } {\bf 4} 187-99 

\bibitem{Lamalle:PPCF1997} 
Lamalle P.U. 1997 {\em   Plasma Phys. Control. Fusion } {\bf 39} 1409-60 

\bibitem{Eester:PPCF2005}
Dirk Van Eester 2005  {\em   Plasma Phys. Control. Fusion } {\bf 47 } 459?481
\bibitem{Johnson:NF2006} 
T. Johnson and T. Hellsten and L.-G. Eriksson 2006 {\em   Nuclear Fusion  } {\bf 46} S433 

\bibitem{Harvey:RF2001}
R. Harvey, Y. Petrov, E. Jaeger, and R. Group 2011 {\em 19th Topical Conference on Radio Frequency Power in Plasmas } {\bf 1406} 369.


\bibitem{Kaufman:PF1972}
Kaufman, A.N., 1972 {\em  The Physics of Fluids} {\bf  15}1063-1069.
  \bibitem{Eriksson:PoP1994}
  Eriksson, L.G. and Helander, P., 1994  {\em Physics of plasmas } {\bf  1} 308-314.
  
    \bibitem{Eriksson:PoP2005}
 Eriksson, L.G. and Schneider, M. 2005 {\em Physics of plasmas } {\bf  12} 072524.

 
 \bibitem{smithe1989local}
D. Smithe 1989  {\em Plasma Physics and Controlled Fusion} {\bf 31} 1105
    \bibitem{Harvey:IAEA1992}
  R. W. Harvey, and M. G. McCoy 1992  {\em Proc. IAEA TCM on Advances in Sim. and Modeling of Thermonuclear Plasmas} USDOC/NTIS No. DE93002962   489-526.
  
  
\bibitem{Lee:PoP2017}
J. P. Lee, J. C. Wright, N. Bertelli, E. F. Jaeger, E. Valeo, R. Harvey, and P. T. Bonoli,  2017 {\em Physics of Plasmas } {\bf 24} 052502
  
\bibitem{Bertelli:NF2017}
N. Bertelli, E.J. Valeo, D.L. Green, M. Gorelenkova, C.K. Phillips, M. Podest{\'a}, J.P. Lee, J.C. Wright and E.F. Jaeger 2017 {\em Nuclear Fusion } {\bf 57} 056035  
 
  \bibitem{Jaeger:NF2006}
E. F. Jaeger, R. Harvey, L. A. Berry, J. Myra, R. Dumont C. Phillips, D. Smithe, R. Barrett, D. Batchelor, P. Bonoli, M. Carter, E. D?azevedo, D. D?ippolito, R. Moore, and J. Wright 2006 {\em Nuclear Fusion} {\bf 46} S397
 
    \bibitem{Levey:RS1969}
  L. Levey and L. B. Felsen 1969 {\em Radio Science} {\bf 4} 959-969
  \bibitem{Cwik:RS1988}
  T. Cwik 1988 {\em Radio Science} {\bf 23} 1133-1140
    \bibitem{budny2012benchmarking}
  R. Budny, L. Berry, R. Bilato, P. Bonoli, M. Brambilla, R.J. Dumont, A. Fukuyama, R. Harvey, E.F. Jaeger, K. Indireshkumar, E. Lerche, D. McCune, C.K. Phillips, V. Vdovin, J. Wright, and, members of the ITPA-IOS, 2012 {\em Nuclear Fusion} {\bf 52|} 023023

  \bibitem{Jucker:CPC2011}
M. Jucker, J. P. Graves, W. A. Cooper, N. Mellet, T. Johnson, and S. Brunner 2011 {\em Computer Physics Communications} {\bf 182} 912

 \bibitem{Hellsten:2004}
T. Hellsten, T. Johnson, J. Carlsson, L.-G. Eriksson, J. Hedin, M. Laxåback and M. Mantsinen 2004 {\em Nuclear Fusion } {\bf  44 }892
 \bibitem{Hellsten:1995}
T. Hellsten, J. Carlsson, L.-G. Eriksson 1995 {\em Physics Review Letters} {\b 74 } 3612
 \bibitem{Murakami:2016}
S. Murakami, , K. Itoh, L. J. Zheng, J. W. Van Dam, P. Bonoli, J. E. Rice, C. L. Fiore, C. Gao, and A. Fukuyama 2016 {\em Physics of plasmas } {\bf  23} 012501
  \bibitem{Kasilov:NF1990}
  S. V. Kasilov, A. I. Pyatak, and K. N. Stepanov 1990  {\em Nuclear Fusion } {\bf 30} 2467  
\bibitem{Jaeger:NF1988}
E. F. Jaeger, D. B. Batchelor, and H. Weitzner 1988  {\em Nuclear Fusion } {\bf 28} 53
\bibitem{Brambilla:1988PPCF}
M. Brambilla and T. Kruchen 1988  {\em Plasma Physics and Controlled Fusion} {\bf 30} 1083 


  \end{thebibliography}
 \end{document}